\documentclass[preprint,12pt]{elsarticle}

\usepackage{amssymb}
\usepackage{amsthm}

\usepackage{float}
\usepackage{color}
\usepackage{amsmath}

\usepackage{graphicx, subfigure}
\usepackage{pslatex}
\usepackage{relsize}
\usepackage{bm}
\usepackage{verbatim}

\usepackage{booktabs}
\usepackage{xcolor}
\usepackage{soul}

\usepackage[utf8]{inputenc}
\usepackage[T1]{fontenc}

\biboptions{numbers,sort&compress}

\newcounter{bla}

\journal{Computer Physics Communications}

\begin{document}

\begin{frontmatter}



\title{MAELAS: MAgneto-ELAStic properties calculation via computational high-throughput approach}


\author[a]{P.~Nieves\corref{author}}
\author[a]{S.~Arapan}
\author[b,c]{S.~H.~Zhang}
\author[a]{A.~P.~K\k{a}dzielawa}
\author[b,c]{R.~F.~Zhang }
\author[a]{D.~Legut}

\cortext[author] {Corresponding author.\\\textit{E-mail address:} pablo.nieves.cordones@vsb.cz}
\address[a]{IT4Innovations, V\v{S}B - Technical University of Ostrava, 17. listopadu 2172/15, 70800 Ostrava-Poruba, Czech Republic}
\address[b]{School of Materials Science and Engineering, Beihang University, Beijing 100191, PR China}
\address[c]{Center for Integrated Computational Materials Engineering, International Research Institute for Multidisciplinary Science, Beihang University, Beijing
100191, PR China}

\begin{abstract}

In this work, we present the program MAELAS to calculate magnetocrystalline anisotropy energy, anisotropic magnetostrictive coefficients and magnetoelastic constants in an automated way by Density Functional Theory calculations. The program is based on the length optimization of the unit cell proposed by Wu and Freeman to calculate the magnetostrictive coefficients for cubic crystals. In addition to cubic crystals, this method is also implemented and generalized for other types of crystals that may be of interest in the study of magnetostrictive materials. As a benchmark, some tests are shown for well-known magnetic materials. 

\end{abstract}

\begin{keyword}
Magnetostriction \sep Magnetoelasticity \sep High-throughput computation \sep First-principles calculations

\end{keyword}

\end{frontmatter}



{\bf PROGRAM SUMMARY}

\begin{small}
\noindent
{\em Program Title:} MAELAS \\
{\em Developer's respository link:} https://github.com/pnieves2019/MAELAS \\
{\em Licensing provisions:} BSD 3-clause \\
{\em Programming language:} Python3 \\
{\em Nature of problem:} To calculate anisotropic magnetostrictive coefficients and magnetoelastic constants in an automated way based on Density Functional Theory methods.\\
{\em Solution method:} In the first stage, the unit cell is relaxed through a spin-polarized calculation without SOC.  Next, after a crystal symmetry analysis, a set of deformed lattice and spin configurations are  generated using the pymatgen library \cite{pymatgenlib}. The energy of these states is calculated by the first-principles code VASP \cite{VASPcode}, including the SOC. The anisotropic magnetostrictive coefficients are derived from the fitting of these energies to a quadratic polynomial \cite{Wu_Freeman}. Finally, if the elastic tensor is provided \cite{AELAScode}, then the magnetoelastic constants are calculated too. \\
{\em Additional comments including restrictions and unusual features:} This version supports the following crystal systems: Cubic (point groups $432$, $\bar{4}3m$, $m\bar{3}m$), Hexagonal ($6mm$, $622$, $\bar{6}2m$, $6/mmm$), Trigonal ($32$, $3m$, $\bar{3}m$), Tetragonal ($4mm$, $422$, $\bar{4}2m$, $4/mmm$) and Orthorhombic ($222$, $2mm$, $mmm$).
\\
   \\

\end{small}

\section{Introduction}
\label{section:Intro}

A magnetostrictive material is one which changes in size due to a change of state of magnetization. These materials are characterized by magnetostrictive coefficients ($\lambda$). In many technical applications such as electric transformers, motor shielding, and magnetic recording, magnetic materials with extremely small magnetostrictive coefficients are required. By contrast, materials with large magnetostrictive coefficients are needed for many applications in electromagnetic microdevices as actuators and sensors \cite{Gibbs,sensors_actuators,sensors,actuator}. Typically, elementary Rare-Earth (R) metals (under low temperature and high magnetic field) and compounds with R and transition metals exhibit a high magnetostriction ($\lambda>10^{-3}$). In particular, the highest
magnetostrictions were found in the RFe$_2$ compounds with Laves phase C15 structure type (face centered cubic) \cite{CLARK1980531}. For instance, Terfenol-D (Tb$_{0.27}$Dy$_{0.73}$Fe$_2$) is a widely used magnetostrictive material thanks to its giant magnetostriction along [111] crystallographic direction ($\lambda_{111}=1.6\times10^{-3}$) under moderate magnetic fields ($<2$ kOe) at room temperature \cite{Eng}. Beyond cubic systems, the research of magnetostrictive materials has been focused on hexagonal crystals like RCo$_5$ (space group 191), hexagonal and trigonal R$_2$Co$_7$ and R$_2$Co$_{17}$ series, and tetragonal R$_2$Fe$_{14}$B \cite{ANDREEV199559,Cullen}. More recently, the problem of R availability \cite{MASSARI201336} has also motivated the exploration of R-free magnetostrictive materials like Galfenol (Fe-Ga), spinel ferrites (CoFe$_2$O$_4$), Nitinol (Ni-Ti alloys), Fe-based Invars, and Ni$_2$MnGa \cite{Frit,Wang2013,Dapino}. 

Concerning the theory of magnetostriction, the basic equations  for cubic (I) crystals were developed by Akulov \cite{Akulov} and Becker et al. \cite{Becker} in the 1920s and 30s. In the next three decades, great advances took place due to the outstanding works of Mason \cite{Mason}, Clark et al. \cite{Clark}, and Callen and Callen \cite{Callen}, as well as many others, where the theory was extended to other crystal symmetries. Over the last decades, modern electronic structure theory based on Density Functional Theory (DFT) has been successfully applied to describe magnetostriction of many materials \cite{Gibbs,Wu1996,WU1997,Wu,Burkert,Pet,Zhang2010,Zhang2011,Zhang2012,Hong,Gav,Wu1999,Wang2013,Frit,FeNi}. Nowadays, a common method to calculate magnetostrictive coefficients is based on the optimization of the unit cell length proposed by Wu and Freeman for cubic crystals  \cite{Wu1996,WU1997}. In this work, we present the MAELAS program where this methodology is implemented and generalized for the main crystal symmetries in the research field of magnetostriction. The paper is organized as follows. In Section \ref{section:theory}, we review some theoretical concepts and equations of magnetostriction. In Section \ref{section:methodology}, we explain in detail the methodology and workflow of the program, while some examples are shown in Section \ref{section:tests}. The paper ends with a summary of the main conclusions and future perspectives (Section \ref{section:conclusion}).

\section{Theory of magnetostriction}
\label{section:theory}

The magnetostrictive response is mainly originated by two kind of sources: (i) isotropic exchange interaction and (ii) strain dependence of magnetocrystalline anisotropy  \cite{Cullen}. The magnetostriction due to isotropic exchange leads to fractional volume changes, and doesn't depend on the magnetization direction \cite{WASSERMAN1990237}. On the other hand, the strain dependence of magnetocrystalline anisotropy is responsible for the magnetostriction that depends on the magnetization orientation (anisotropic), and is originated by the spin-orbit coupling (SOC) and crystal field interactions \cite{Cullen,Skomskibook}. The current version of the program MAELAS calculates the magnetostrictive coefficients and magnetoelastic constants related to the anisotropic magnetostriction.

\begin{figure}[h!]
\centering
\includegraphics[width=\columnwidth ,angle=0]{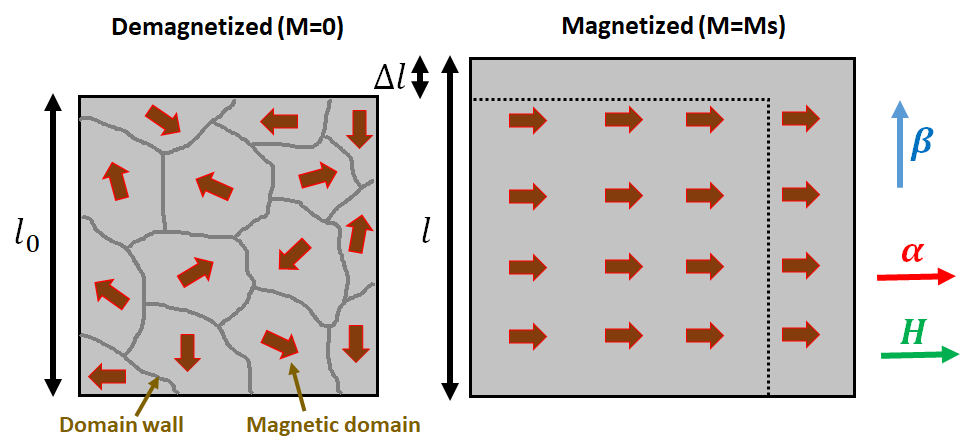}
\caption{Magnetostriction of a single crystal under an external magnetic field ($\boldsymbol{\alpha}\|\boldsymbol{H}$) perpendicular to the measured length direction ($\boldsymbol{\beta}\bot\boldsymbol{H}$). Symbols $M$ and $M_s$ stand for macroscopic magnetization and saturation magnetization, respectively. Dash line on the right represents the original size of the demagnetized material. The magnetostriction effect has been magnified in order to help to visualize it easily, in real materials it is smaller ($\Delta l/l_0\sim10^{-3}-10^{-6}$). }
\label{fig:diagram}
\end{figure}

Let's consider $l_0$ the initial length of a demagnetized material along the direction $\boldsymbol{\beta}$ ($\vert \boldsymbol{\beta}\vert=1$), and $l$ the final length along the same direction $\boldsymbol{\beta}$ when the system is magnetized along the direction $\boldsymbol{\alpha}$ ($\vert \boldsymbol{\alpha}\vert=1$). The relative length change $(l-l_0)/l_0=\Delta l/l_0$ can be written as \cite{Cullen}
\begin{equation}
    \frac{\Delta l}{l_0}\Bigg\vert_{\boldsymbol{\beta}}^{\boldsymbol{\alpha}}=\sum_{i,j=x,y,z}\epsilon_{ij}^{eq}(\boldsymbol{\alpha})\beta_i\beta_j,
    \label{eq:delta_l}
\end{equation}
where $\epsilon_{ij}^{eq}$ is the equilibrium strain tensor. This equation describes the Joule effect \cite{Joule}, once it is rewritten in terms of the magnetostrictive coefficients ($\lambda$) conveniently. Fig.\ref{fig:diagram} shows a sketch of magnetostriction. 

 The  deformation of a solid can be described in terms of the displacement vector $\boldsymbol{u}(\boldsymbol{r})=\boldsymbol{r'}-\boldsymbol{r}$ that gives the displacement of a point at the initial position $\boldsymbol{r}$  to its final position $\boldsymbol{r'}$ after it is deformed.  For small deformations (infinitesimal strain theory), the strain tensor ($\epsilon_{ij}$) can be expressed in terms of the displacement vector as\cite{Landau}
\begin{equation}
\begin{aligned}
     \epsilon_{ij}=\frac{1}{2}\left(\frac{\partial u_{i}}{\partial r_{j}}+\frac{\partial u_j}{\partial r_i}\right),\quad\quad i,j=x,y,z
    \label{eq:disp_vec}
\end{aligned}
\end{equation}
where $\partial u_i/ \partial r_j$ is called the displacement gradient (second-order tensor). The equilibrium strain tensor is obtained through the minimization of both the elastic ($E_{el}$) and magnetoelastic ($E_{me}$) energies \cite{CLARK1980531,Cullen}
\begin{equation}
    \frac{\partial(E_{el}+E_{me})}{\partial\epsilon_{ij}}=0, \quad\quad i,j=x,y,z
    \label{eq:dE}
\end{equation}
where the total energy must be invariant under the symmetry operations of the crystal lattice \cite{CLARK1980531}. Let's write general equations for $E_{el}$ and $E_{me}$. The elastic energy depends on the fourth-order elastic stiffness tensor $c_{ijkl}$ that links the second-order strain and stress ($\sigma_{ij}$) tensors through the generalized Hooke's law
\begin{equation}
    \sigma_{ij} = \sum_{k,l=x,y,z}c_{ijkl}\epsilon_{kl}, \quad i,j=x,y,z.
    \label{eq:hooke}
\end{equation}
Taking advantage of the symmetry of stress and strain tensors, the Hooke's law can be written in matrix notation as
\begin{equation}
\begin{aligned}
    \begin{pmatrix}
\sigma_{xx} \\
\sigma_{yy}   \\
\sigma_{zz}   \\
\sigma_{yz}   \\
\sigma_{xz}   \\
\sigma_{xy}   \\
\end{pmatrix} & =\begin{pmatrix}
c_{xxxx} & c_{xxyy} & c_{xxzz} & c_{xxyz} & c_{xxzx} & c_{xxxy} \\
c_{yyxx} & c_{yyyy} & c_{yyzz} & c_{yyyz} & c_{yyzx} & c_{yyxy} \\
c_{zzxx} & c_{zzyy} & c_{zzzz} & c_{zzyz} & c_{zzzx} & c_{zzxy} \\
c_{yzxx} & c_{yzyy} & c_{yzzz} & c_{yzyz} & c_{yzzx} & c_{yzxy} \\
c_{zxxx} & c_{zxyy} & c_{zxzz} & c_{zxyz} & c_{zxzx} & c_{zxxy} \\
c_{xyxx} & c_{xyyy} & c_{xyzz} & c_{xyyz} & c_{xyzx} & c_{xyxy} \\
\end{pmatrix}
    \begin{pmatrix}
\epsilon_{xx} \\
\epsilon_{yy}   \\
\epsilon_{zz}   \\
2\epsilon_{yz}\\
2\epsilon_{xz} \\
2\epsilon_{xy} \\
\end{pmatrix}
\label{eq:stiffnees_tensor}
\end{aligned}
\end{equation}
To facilitate the manipulation of this equation it is convenient to define the following six-dimensional vectors (Voigt notation)
\begin{equation}
\begin{aligned}
  \boldsymbol{\tilde{\sigma}} =\begin{pmatrix}
\tilde{\sigma}_{1} \\
\tilde{\sigma}_{2}   \\
\tilde{\sigma}_{3}   \\
\tilde{\sigma}_{4}   \\
\tilde{\sigma}_{5}   \\
\tilde{\sigma}_{6}   \\
\end{pmatrix} 
=\begin{pmatrix}
\sigma_{xx} \\
\sigma_{yy}   \\
\sigma_{zz}   \\
\sigma_{yz}   \\
\sigma_{xz}   \\
\sigma_{xy}   \\
\end{pmatrix}, \quad\quad
 \boldsymbol{\tilde{\epsilon}} =\begin{pmatrix}
\tilde{\epsilon}_{1} \\
\tilde{\epsilon}_{2}   \\
\tilde{\epsilon}_{3}   \\
\tilde{\epsilon}_{4}   \\
\tilde{\epsilon}_{5}   \\
\tilde{\epsilon}_{6}   \\
\end{pmatrix} 
=\begin{pmatrix}
\epsilon_{xx} \\
\epsilon_{yy}   \\
\epsilon_{zz}   \\
2\epsilon_{yz}   \\
2\epsilon_{xz}   \\
2\epsilon_{xy}   \\
\end{pmatrix},
\label{eq:vector_stress_strain}
\end{aligned}
\end{equation}
and replace $c_{ijkl}$ by $C_{nm}$ contracting a pair of cartesian indices into a single integer: $xx\rightarrow1$, $yy\rightarrow2$, $zz\rightarrow3$, $yz\rightarrow4$, $xz\rightarrow5$ and $xy\rightarrow6$. Using these conversion rules the Hooke's law is simplified to 
\begin{equation}
    \tilde{\sigma}_{i} = \sum_{j=1}^6 C_{ij}\tilde{\epsilon}_{j}, \quad i=1,...,6
    \label{eq:hooke_simple}
\end{equation}
where in matrix form reads
\begin{equation}
\begin{aligned}
    \begin{pmatrix}
\tilde{\sigma}_{1} \\
\tilde{\sigma}_{2}   \\
\tilde{\sigma}_{3}   \\
\tilde{\sigma}_{4}   \\
\tilde{\sigma}_{5}   \\
\tilde{\sigma}_{6}   \\
\end{pmatrix} & =\begin{pmatrix}
C_{11} & C_{12} & C_{13} & C_{14} & C_{15} & C_{16} \\
C_{21} & C_{22} & C_{23} & C_{24} & C_{25} & C_{26} \\
C_{31} & C_{32} & C_{33} & C_{34} & C_{35} & C_{36} \\
C_{41} & C_{42} & C_{42} & C_{44} & C_{45} & C_{46} \\
C_{51} & C_{52} & C_{53} & C_{54} & C_{55} & C_{56} \\
C_{61} & C_{62} & C_{63} & C_{64} & C_{65} & C_{66} \\
\end{pmatrix}
    \begin{pmatrix}
\tilde{\epsilon}_{1} \\
\tilde{\epsilon}_{2}   \\
\tilde{\epsilon}_{3}   \\
\tilde{\epsilon}_{4}   \\
\tilde{\epsilon}_{5}   \\
\tilde{\epsilon}_{6}   \\
\end{pmatrix}.
\label{eq:stiffnees_tensor_voigt}
\end{aligned}
\end{equation}
where $C_{ij}=C_{ji}$. Then the elastic energy up to second-order in the strain can be written as
\begin{equation}
    E_{el}=E_0+\frac{V_0}{2}\sum_{i,j=1}^6C_{ij}\tilde{\epsilon}_i\tilde{\epsilon}_j+O (\tilde{\epsilon}^3),
    \label{eq:E_el}
\end{equation}
where $E_0$ and $V_0$ are the equilibrium energy and volume, respectively. The magnetoelastic energy $E_{me}$ comes from the strain dependence of the magnetocrystalline anisotropy energy (MAE) $E_K$ \cite{Birss,kittel1949}. Performing a Taylor expansion of $E_K$ in the strain we have
\begin{equation}
    E_{K}  = E_{K}^0+\sum_{i=1}^{6}\left(\frac{\partial E_K}{\partial \tilde{\epsilon}_i }\right)_0\tilde{\epsilon}_i+\frac{1}{2}\sum_{i,j=1}^{6}\left(\frac{\partial^2 E_K}{\partial \tilde{\epsilon}_i \partial\tilde{\epsilon}_j}\right)_0\tilde{\epsilon}_i\tilde{\epsilon}_j+O (\tilde{\epsilon}^3),
    \label{eq:E_K}
\end{equation}
where $E_{K}^0$ corresponds to the MAE of the undeformed state that contains the magnetocrystalline anisotropy constants $K$. The third term in the right hand side of Eq.\ref{eq:E_K} is the second-order magnetoelastic energy that leads to a very small additional contribution to the second-order elastic energy given by Eq.\ref{eq:E_el}, so that is usually neglected \cite{Birss,Mueller}. The first-order magnetoelastic energy 
\begin{equation}
    E_{me}  =\sum_{i=1}^{6}\left(\frac{\partial E_K}{\partial \tilde{\epsilon}_i }\right)_0\tilde{\epsilon}_i
    \label{eq:E_me_0}
\end{equation}
is obtained by taking the direct product of the symmetry strains and direction cosine polynomial for each irreducible representation, multiplying by a constant, called the magnetoelastic constant and finally summing over the different representations \cite{Callen,Clark,CLARK1980531,Cullen}. Frequently, the first-order magnetoelastic energy is considered up to second-order of the direction cosine polynomial $\alpha$. In cartesian coordinates, it may be written as
\begin{equation}
    E_{me}=\sum_{i=1}^3g_i(\alpha^0)\tilde{\epsilon}_i+\sum_{i=1}^{6}f_i(\alpha^2)\tilde{\epsilon}_i+O(\alpha^4),
    \label{eq:E_me}
\end{equation}
where functions $g_i$ and $f_i$ contain the magnetoelastic constants ($b$).  In the following subsections, we show the form of Eqs.\ref{eq:delta_l}, \ref{eq:E_el} and \ref{eq:E_me} for the main crystal symmetries studied in magnetostriction, which are implemented in the program MAELAS. The remaining crystal systems not discussed here might be included in the new versions of the code. In Table \ref{tab:maelas_systems}, we present a summary of the crystal systems supported by MAELAS. Here, we use the notation of Wallace \cite{Wallace,mouhat} (I/II) to distinguish Laue classes within the same crystal system. 

Before analyzing each crystal system, we must make an important remark about the notation for the strain tensor $\epsilon_{ij}$. In previous works discussing magnetostriction like  Refs.\cite{kittel1949,CLARK1980531,Cullen}, the Voigt definition of the strain tensor was used ($\epsilon^{V}_{ij}, i,j=x,y,z$) \cite{Love}, which is related to the one defined in the present work as $\epsilon_{ii}=\epsilon_{ii}^V$, $2\epsilon_{ij}=\epsilon_{ij}^V, i\neq j$. Consequently, the following elastic and magnetoelastic energies (in terms of the strain tensor with two cartesian indices) contain numerical factors different to those given in Kittel and Clark  works \cite{kittel1949,CLARK1980531} for the terms with non-diagonal elements of the strain tensor ($\epsilon_{ij}, i\neq j$). The following expressions for the relative length change ($\Delta l/l_0$) are the same as in Kittel and Clark works \cite{kittel1949,CLARK1980531} because the sum in Eq.\ref{eq:delta_l} runs over all possible values of indices $i,j=x,y,z$, while in Kittel and Clark works \cite{kittel1949,CLARK1980531} the sum runs up to $i\geqslant j$. In the present work, the equations of magnetostrictive coefficients expressed in terms of the elastic and magnetoelastic constants are also the same to those given in Kittel and Clark works \cite{kittel1949,CLARK1980531}.

\begin{table}[]
\centering
\caption{Number of independent second-order elastic constants of each crystal system. Number of independent magnetoelastic and magnetostrictive coefficients up to second-order of the direction cosine polynomial in the first-order magnetoelastic energy. In the last column we specify which crystal systems are supported by the current version of MAELAS. }
\label{tab:maelas_systems}
\resizebox{\textwidth}{!}{%
\begin{tabular}{@{}ccccccc@{}}
\toprule
Crystal system &
  Point groups &
  \begin{tabular}[c]{@{}c@{}}Space \\ groups\end{tabular} &
  \begin{tabular}[c]{@{}c@{}}Elastic \\ constants \\ ($C_{ij}$)\end{tabular} &
  \begin{tabular}[c]{@{}c@{}}Magnetoelastic \\ constants \\ ($b$)\end{tabular} &
  \begin{tabular}[c]{@{}c@{}}Magnetostriction \\ coefficients \\ ($\lambda$)\end{tabular} &
  MAELAS  \\ \midrule  \hline
Triclinic       & $1,\bar{1}$               & $1-2$     & 21 & 36 & 36 & No  \\
Monoclinic      & $2,m,2/m$                 & $3-15$    & 13 & 20 & 20 & No  \\
Orthorhombic    & $222,2mm,mmm$             & $16-74$   & 9  & 12 & 12 & Yes \\
Tetragonal (II) & $4,\bar{4}, 4/m$          & $75-88$   & 7  & 10 & 10 & No  \\
Tetragonal (I)  & $4mm,422,\bar{4}2m,4/mmm$ & $89-142$  & 6  & 7  & 7  & Yes \\
Trigonal (II)   & $3,\bar{3}$               & $143-148$ & 7  & 12 & 12 & No  \\
Trigonal (I)    & $32,3m,\bar{3}m$          & $149-167$ & 6  & 8  & 8  & Yes \\
Hexagonal(II)   & $6,\bar{6},6/m$           & $168-176$ & 5  & 8  & 8  & No  \\
Hexagonal (I)   & $6mm,622,\bar{6}2m,6/mmm$ & $177-194$ & 5  & 6  & 6  & Yes \\
Cubic (II)      & $23,m\bar{3}$             & $195-206$ & 3  & 4  & 4  & No  \\
Cubic (I)       & $432,\bar{4}3m,m\bar{3}m$ & $207-230$ & 3  & 3  & 3  & Yes \\ \bottomrule
\end{tabular}%
}
\end{table}

\subsection{Cubic (I)}
\label{subsection:cubic_I}

\subsubsection{Single crystal}

For cubic (I) systems (point groups $432$, $\bar{4}3m$, $m\bar{3}m$) the elastic stiffness tensor reads
\begin{equation}
\begin{aligned}
C^{cub} =\begin{pmatrix}
C_{11} & C_{12} & C_{12} & 0 & 0 & 0 \\
C_{12} & C_{11} &  C_{12}& 0 & 0 & 0 \\
C_{12} & C_{12} & C_{11} & 0& 0 & 0 \\
0 & 0 & 0 & C_{44} & 0 & 0 \\
0 & 0 & 0 & 0 & C_{44} & 0 \\
0 & 0 & 0 & 0 & 0 & C_{44} \\
\end{pmatrix},
\label{eq:stiffnees_tensor_cub}
\end{aligned}
\end{equation}
so there are three independent elastic constants $C_{11}$, $C_{12}$ and $C_{44}$. Hence, the elastic energy Eq.\ref{eq:E_el} becomes
\begin{equation}
\begin{aligned}
\frac{E_{el}^{cub}-E_0}{V_0} & =\frac{C_{11}}{2}(\tilde{\epsilon}_{1}^2+\tilde{\epsilon}_{2}^2+\tilde{\epsilon}_{3}^2)+C_{12}(\tilde{\epsilon}_{1}\tilde{\epsilon}_{2}+\tilde{\epsilon}_{1}\tilde{\epsilon}_{3}+\tilde{\epsilon}_{2}\tilde{\epsilon}_{3})\\
& + \frac{C_{44}}{2}(\tilde{\epsilon}_{4}^2+\tilde{\epsilon}_{5}^2+\tilde{\epsilon}_{6}^2)\\
& = \frac{c_{xxxx}}{2}(\epsilon_{xx}^2+\epsilon_{yy}^2+\epsilon_{zz}^2)+c_{xxyy}(\epsilon_{xx}\epsilon_{yy}+\epsilon_{xx}\epsilon_{zz}+\epsilon_{yy}\epsilon_{zz})\\
& + 2c_{yzyz}(\epsilon_{xy}^2+\epsilon_{yz}^2+\epsilon_{xz}^2),
\end{aligned}
\label{eq:E_el_cub}
\end{equation}
where $C_{11}=c_{xxxx}$, $C_{12}=c_{xxyy}$ and $C_{44}=c_{yzyz}$. On the other hand, the first-order magnetoelastic energy up to second-order direction cosine polynomial contains 3 magnetoelastic constants \cite{Callen}. From the symmetry strains and direction cosine polynomial for each irreducible representation, it is possible to obtain the following magnetoelastic energy in cartesian coordinates \cite{CLARK1980531,Cullen,Frit}
\begin{equation}
\begin{aligned}
    \frac{E_{me}^{cub(I)}}{V_0}  & =  b_0(\tilde{\epsilon}_{1}+\tilde{\epsilon}_{2}+\tilde{\epsilon}_{3})+b_1(\alpha_x^2\tilde{\epsilon}_{1}+\alpha_y^2\tilde{\epsilon}_{2}+\alpha_z^2\tilde{\epsilon}_{3})\\
    & +  b_2(\alpha_x\alpha_y\tilde{\epsilon}_{6}+\alpha_x\alpha_z\tilde{\epsilon}_{5}+\alpha_y\alpha_z\tilde{\epsilon}_{4}) \\
    & =b_0(\epsilon_{xx}+\epsilon_{yy}+\epsilon_{zz})+b_1(\alpha_x^2\epsilon_{xx}+\alpha_y^2\epsilon_{yy}+\alpha_z^2\epsilon_{zz})\\
    & +  2b_2(\alpha_x\alpha_y\epsilon_{xy}+\alpha_x\alpha_z\epsilon_{xz}+\alpha_y\alpha_z\epsilon_{yz}), 
\label{eq:E_me_cub_I}     
\end{aligned}
\end{equation}
where $b_0$ is the volume magnetoelastic constant, and $b_1$ and $b_2$ are the anisotropic magnetoelastic constants. Next, replacing Eqs.\ref{eq:E_el_cub} and \ref{eq:E_me_cub_I} into Eq.\ref{eq:dE}, we find the following equilibrium strains
\begin{equation}
    \begin{aligned}
        \epsilon_{ij}^{eq} & =  -\frac{b_2\alpha_{i}\alpha_{j}}{2C_{44}},\quad\quad i\neq j,\quad\quad i,j=x,y,z\\
        \epsilon_{ii}^{eq} & =  -\frac{b_1\alpha_i^2}{C_{11}-C_{12}}-\frac{b_0}{C_{11}+2C_{12}}+\frac{b_1C_{12}}{(C_{11}-C_{12})(C_{11}+2C_{12})}, \quad i=x,y,z
    \end{aligned}
\end{equation}
Inserting these equilibrium strains into Eq.\ref{eq:delta_l} gives
\begin{equation}
\begin{aligned}
     \frac{\Delta l}{l_0}\Bigg\vert_{\boldsymbol{\beta}}^{\boldsymbol{\alpha}} & =\lambda^\alpha+\frac{3}{2}\lambda_{001}\left(\alpha_x^2\beta_{x}^2+\alpha_y^2\beta_{y}^2+\alpha_z^2\beta_{z}^2-\frac{1}{3}\right)\\
     & + 3\lambda_{111}(\alpha_x\alpha_y\beta_{x}\beta_{y}+\alpha_y\alpha_z\beta_{y}\beta_{z}+\alpha_x\alpha_z\beta_{x}\beta_{z}),
    \label{eq:delta_l_cub_I}
\end{aligned}
\end{equation}
where 
\begin{equation}
    \begin{aligned}
        \lambda^\alpha & = \frac{-b_0-\frac{1}{3}b_1}{C_{11}+2C_{12}},\\
        \lambda_{001} & = \frac{-2b_1}{3(C_{11}-C_{12})},\\
        \lambda_{111} & = \frac{-b_2}{3C_{44}}.
    \label{eq:lamb_cub}
    \end{aligned}
\end{equation}
The coefficient $\lambda^\alpha$ describes the
volume magnetostriction, while $\lambda_{001}$ and $\lambda_{111}$ are the anisotropic magnetostrictive coefficients that give the fractional length
change along the [001] and [111] directions when a demagnetized material is magnetized in these directions, respectively. The superscript $\alpha$ in $\lambda^\alpha$ stands for one irreducible representation of the group of
transformations which take the crystal into
itself \cite{CLARK1980531,Cullen}, so it should not be confused with the direction of magnetization $\boldsymbol{\alpha}$. The MAE in an unstrained cubic crystal up to sixth-order of direction cosine polynomial is \cite{Handley,kittel1949}
\begin{equation}
    \frac{E_{K}^0}{V_0}= K_0+K_1(\alpha_x^2\alpha_y^2+\alpha_x^2\alpha_z^2+\alpha_y^2\alpha_z^2)+K_2\alpha_x^2\alpha_y^2\alpha_z^2, 
\label{eq:E_mca_cub}     
\end{equation}
where $K_0$, $K_1$ and $K_2$ are the magnetocrystalline anisotropy constants.

\subsubsection{Polycrystal}

The theory of magnetostriction for polycrystalline materials is more complex than for single crystals. A widely used approximation is to assume that the stress distribution is uniform through the material. In this case the relative change in length may be put into the form  \cite{Akulov,Lee_1955,Cullen,Birss}
\begin{equation}
     \frac{\Delta l}{l_0}\Bigg\vert_{\boldsymbol{\beta}}^{\boldsymbol{\alpha}} = \frac{3}{2}\lambda_{S}\left[(\boldsymbol{\alpha}\cdot\boldsymbol{\beta})^2-\frac{1}{3}\right],
    \label{eq:delta_l_cub_poly}
\end{equation}
where
\begin{equation}
\lambda_S=\frac{2}{5}\lambda_{001}+\frac{3}{5}\lambda_{111}.
\end{equation}
This result is analogous to the Reuss approximation used in the elastic theory of polycrystals to obtain a lower bound of bulk and shear modulus \cite{Cullen,Reuss,Hill_1952,AELAS}. A discussion about the limitations of this approximation can be found in Ref.\cite{tremolet}.

\subsection{Hexagonal (I)}
\label{subsection:hexagonal_I}

\subsubsection{Single crystal}

The elastic stiffness tensor for hexagonal (I) system (point groups $6mm$, $622$, $\bar{6}2m$, $6/mmm$) reads
\begin{equation}
\begin{aligned}
C^{hex} =\begin{pmatrix}
C_{11} & C_{12} & C_{13} & 0 & 0 & 0 \\
C_{12} & C_{11} &  C_{13}& 0 & 0 & 0 \\
C_{13} & C_{13} & C_{33} & 0& 0 & 0 \\
0 & 0 & 0 & C_{44} & 0 & 0 \\
0 & 0 & 0 & 0 & C_{44} & 0 \\
0 & 0 & 0 & 0 & 0 & \frac{C_{11}-C_{12}}{2} \\
\end{pmatrix},
\label{eq:stiffnees_tensor_hex}
\end{aligned}
\end{equation}
so that it has five independent elastic constants $C_{11}$, $C_{12}$, $C_{13}$, $C_{33}$ and $C_{44}$. As a result, the elastic energy Eq.\ref{eq:E_el} is
\begin{equation}
\begin{aligned}
 \frac{ E_{el}^{hex}-E_0}{V_0} & = \frac{1}{2}C_{11}(\tilde{\epsilon}_{1}^2+\tilde{\epsilon}_{2}^2)+C_{12}\tilde{\epsilon}_{1}\tilde{\epsilon}_{2}+C_{13}(\tilde{\epsilon}_{1}+\tilde{\epsilon}_{2})\tilde{\epsilon}_{3}+\frac{1}{2}C_{33}\tilde{\epsilon}_{3}^2 \\
  & +\frac{1}{2}C_{44}(\tilde{\epsilon}_{4}^2+\tilde{\epsilon}_{5}^2)+\frac{1}{4}(C_{11}-C_{12})\tilde{\epsilon}_{6}^2\\
 & = \frac{1}{2}c_{xxxx}(\epsilon_{xx}^2+\epsilon_{yy}^2)+c_{xxyy}\epsilon_{xx}\epsilon_{yy}+c_{xxzz}(\epsilon_{xx}+\epsilon_{yy})\epsilon_{zz}+\frac{1}{2}c_{zzzz}\epsilon_{zz}^2 \\
  & +2c_{yzyz}(\epsilon_{yz}^2+\epsilon_{xz}^2)+(c_{xxxx}-c_{xxyy})\epsilon_{xy}^2
\end{aligned}
\label{eq:E_el_hex}
\end{equation}
where $C_{11}=c_{xxxx}$, $C_{12}=c_{xxyy}$, $C_{13}=c_{xxzz}$, $C_{33}=c_{zzzz}$, and $C_{44}=c_{yzyz}$. The first order magnetoelastic energy up to quadratic direction cosine polynomial contains 6 magnetoelastic constants \cite{Callen}. In cartesian coordinates it can be written as \cite{CLARK1980531}
\begin{equation}
\begin{aligned}
    \frac{E_{me}^{hex(I)}}{V_0}  & =  b_{11}(\epsilon_{xx}+\epsilon_{yy})+b_{12}\epsilon_{zz}+b_{21}\left(\alpha_z^2-\frac{1}{3}\right)(\epsilon_{xx}+\epsilon_{yy})+b_{22}\left(\alpha_z^2-\frac{1}{3}\right)\epsilon_{zz}\\
    & + b_3\left[\frac{1}{2}(\alpha_x^2-\alpha_y^2)(\epsilon_{xx}-\epsilon_{yy})+2\alpha_x\alpha_y\epsilon_{xy}\right]+2b_4(\alpha_x\alpha_z\epsilon_{xz}+\alpha_y\alpha_z\epsilon_{yz}).
\label{eq:E_me_hex_I}     
\end{aligned}
\end{equation}
Once the equilibrium strains are calculated by minimizing Eqs.\ref{eq:E_el_hex} and \ref{eq:E_me_hex_I} through Eq.\ref{eq:dE} and inserted into Eq.\ref{eq:delta_l}, one finds \cite{Clark,CLARK1980531,Cullen}
\begin{equation}
\begin{aligned}
     \frac{\Delta l}{l_0}\Bigg\vert_{\boldsymbol{\beta}}^{\boldsymbol{\alpha}} & =\lambda^{\alpha1,0}(\beta_x^2+\beta_y^2)+\lambda^{\alpha2,0}\beta_z^2+\lambda^{\alpha1,2}\left(\alpha_z^2-\frac{1}{3}\right)(\beta_x^2+\beta_y^2)\\
     & + \lambda^{\alpha2,2}\left(\alpha_z^2-\frac{1}{3}\right)\beta_z^2+\lambda^{\gamma,2}\left[\frac{1}{2}(\alpha_x^2-\alpha_y^2)(\beta_x^2-\beta_y^2)+2\alpha_x\alpha_y\beta_x\beta_y\right]\\
     & + 2\lambda^{\epsilon,2}(\alpha_x\alpha_z\beta_x\beta_z+\alpha_y\alpha_z\beta_y\beta_z),
    \label{eq:delta_l_hex_I}
\end{aligned}
\end{equation}
where
\begin{equation}
    \begin{aligned}
        \lambda^{\alpha1,0} & = \frac{b_{11}C_{33}+b_{12}C_{13}}{C_{33}(C_{11}+C_{12})-2C_{13}^2},\\
        \lambda^{\alpha2,0} & = \frac{2b_{11}C_{13}-b_{12}(C_{11}+C_{12})}{C_{33}(C_{11}+C_{12})-2C_{13}^2},\\
        \lambda^{\alpha1,2} & = \frac{-b_{21}C_{33}+b_{22}C_{13}}{C_{33}(C_{11}+C_{12})-2C_{13}^2},\\
        \lambda^{\alpha2,2} & = \frac{2b_{21}C_{13}-b_{22}(C_{11}+C_{12})}{C_{33}(C_{11}+C_{12})-2C_{13}^2},\\
        \lambda^{\gamma,2} & = \frac{-b_3}{C_{11}-C_{12}},\\
        \lambda^{\epsilon,2} & = \frac{-b_4}{2C_{44}}.
    \label{eq:lamb_hex}
    \end{aligned}
\end{equation}
These magnetostrictive coefficients are related to the normal strain modes for a cylinder \cite{CLARK1980531,Cullen}. The equation of the relative length change in the form of Eq.\ref{eq:delta_l_hex_I} was proposed by Clark et al. \cite{Clark}. In literature there are different arrangements of the right hand side of Eq.\ref{eq:delta_l_hex_I} that leads to other definitions of the magnetostrictive coefficients, like those defined by Mason \cite{Mason}, Birss \cite{Birss}, and Callen and Callen \cite{Callen}. The conversion formulas between Eq.\ref{eq:delta_l_hex_I} and all these other conventions can be found in \ref{app:hex_I}. These conversion formulas are implemented in the program MAELAS, so that the magnetostrictive coefficients are also given according to these definitions. Note that in some works  \cite{tremolet,Handley} the magnetostrictive coefficients $\lambda^{\gamma,2}$ and $\lambda^{\epsilon,2}$ in Eq.\ref{eq:delta_l_hex_I} are named as $\lambda^{\epsilon,2}$ and $\lambda^{\zeta,2}$, respectively, which is more consistent with the Bethe's group-theoretical notation \cite{tremolet}. The MAE in an unstrained hexagonal crystal up to fourth-order of $\alpha$ reads \cite{Handley}
\begin{equation}
\begin{aligned}
    \frac{E_{K}^0}{V_0}= K_0+K_1(1-\alpha_z^2)+K_2(1-\alpha_z^2)^2.
\label{eq:E_mca_hex}     
\end{aligned}
\end{equation}

\subsubsection{Polycrystal}

Under the assumption of uniform stress, the relative change in length for polycrystalline hexagonal (I) systems can be written as \cite{Birss}
\begin{equation}
     \frac{\Delta l}{l_0}\Bigg\vert_{\boldsymbol{\beta}}^{\boldsymbol{\alpha}} = \xi+\eta(\boldsymbol{\alpha}\cdot\boldsymbol{\beta})^2,
    \label{eq:delta_l_hex_poly}
\end{equation}
where $\eta$ is, in both easy axis and easy plane MAE, given by 
\begin{equation}
     \eta=-\frac{2}{15}Q_4+\frac{1}{5}Q_6+\frac{7}{15}Q_8.
    \label{eq:eta_hex_poly}
\end{equation}
The quantity $\xi$ is different for easy axis and easy plane. In the case of easy axis, $\xi$ is given by
\begin{equation}
     \xi=\frac{2}{3}Q_2+\frac{4}{15}Q_4-\frac{1}{15}Q_6+\frac{1}{15}Q_8, \quad\quad \textrm{(easy axis)}
    \label{eq:xi_hex_poly_ea}
\end{equation}
while for easy plane is
\begin{equation}
     \xi=-\frac{1}{3}Q_2-\frac{1}{15}Q_4-\frac{1}{15}Q_6-\frac{4}{15}Q_8. \quad\quad \textrm{(easy plane)}
    \label{eq:xi_hex_poly_ep}
\end{equation}
The quantities $Q_i$ ($i=2,4,6,8$) are the anisotropic magnetostrictive coefficients in Birss's convention \cite{Birss}, and are related to the magnetostrictive coefficients defined in Eq. \ref{eq:delta_l_hex_I} through Eq. \ref{eq:lamb_Birss_hex_I}. We have implemented these formulas in MAELAS, so that it also calculates  $\eta$ and $\xi$.

\subsection{Trigonal (I)}
\label{subsection:trigonal_I}

\subsubsection{Single crystal}

The elastic stiffness tensor for trigonal (I) system (point groups $32$, $3m$, $\bar{3}m$) has 6 independent elastic constants $C_{11}$, $C_{12}$, $C_{13}$, $C_{33}$, $C_{44}$ and $C_{14}$, and it is given by
\begin{equation}
\begin{aligned}
C^{trig(I)} =\begin{pmatrix}
C_{11} & C_{12} & C_{13} & C_{14} & 0 & 0 \\
C_{12} & C_{11} &  C_{13}& -C_{14} & 0 & 0 \\
C_{13} & C_{13} & C_{33} & 0& 0 & 0 \\
C_{14} & -C_{14} & 0 & C_{44} & 0 & 0 \\
0 & 0 & 0 & 0 & C_{44} & C_{14} \\
0 & 0 & 0 & 0 & C_{14} & \frac{C_{11}-C_{12}}{2} \\
\end{pmatrix}.
\label{eq:stiffnees_tensor_trig}
\end{aligned}
\end{equation}
Hence, inserting this tensor into Eq.\ref{eq:E_el} we have the following elastic energy
\begin{equation}
\begin{aligned}
  \frac{E_{el}^{trig(I)}-E_0}{V_0} & = \frac{1}{2}C_{11}(\tilde{\epsilon}_{1}^2+\tilde{\epsilon}_{2}^2)+C_{12}\tilde{\epsilon}_{1}\tilde{\epsilon}_{2}+C_{13}(\tilde{\epsilon}_{1}+\tilde{\epsilon}_{2})\tilde{\epsilon}_{3}+\frac{1}{2}C_{33}\tilde{\epsilon}_{3}^2 \\
  & +\frac{1}{2}C_{44}(\tilde{\epsilon}_{5}^2+\tilde{\epsilon}_{4}^2)+\frac{1}{4}(C_{11}-C_{12})\tilde{\epsilon}_{6}^2+C_{14}(\tilde{\epsilon}_{6}\tilde{\epsilon}_{5}+\tilde{\epsilon}_{1}\tilde{\epsilon}_{4}-\tilde{\epsilon}_{2}\tilde{\epsilon}_{4}).\\
  & = \frac{1}{2}c_{xxxx}(\epsilon_{xx}^2+\epsilon_{yy}^2)+c_{xxyy}\epsilon_{xx}\epsilon_{yy}+c_{xxzz}(\epsilon_{xx}+\epsilon_{yy})\epsilon_{zz}+\frac{1}{2}c_{zzzz}\epsilon_{zz}^2 \\
  & +2c_{yzyz}(\epsilon_{xz}^2+\epsilon_{yz}^2)+(c_{xxxx}-c_{xxyy})\epsilon_{xy}^2+4c_{xxyz}(\epsilon_{xy}\epsilon_{xz}+\epsilon_{xx}\epsilon_{yz}-\epsilon_{yy}\epsilon_{yz}).\\
\end{aligned}
\label{eq:E_el_trig}
\end{equation}
where $C_{11}=c_{xxxx}$, $C_{12}=c_{xxyy}$, $C_{13}=c_{xxzz}$, $C_{14}=c_{xxyz}$, $C_{33}=c_{zzzz}$, and $C_{44}=c_{yzyz}$. On the other hand, the magnetoelastic energy contains 8 independent magnetoelastic constants \cite{Callen}. In cartesian coordinates it can be written as \cite{Cullen}
\begin{equation}
\begin{aligned}
    \frac{E_{me}^{trig(I)}}{V_0}  & =  b_{11}(\epsilon_{xx}+\epsilon_{yy})+b_{12}\epsilon_{zz}+b_{21}\left(\alpha_z^2-\frac{1}{3}\right)(\epsilon_{xx}+\epsilon_{yy})+b_{22}\left(\alpha_z^2-\frac{1}{3}\right)\epsilon_{zz}\\
    & + b_3\left[\frac{1}{2}(\alpha_x^2-\alpha_y^2)(\epsilon_{xx}-\epsilon_{yy})+2\alpha_x\alpha_y\epsilon_{xy}\right]+2b_4(\alpha_x\alpha_z\epsilon_{xz}+\alpha_y\alpha_z\epsilon_{yz})\\
    & + b_{14}\left[(\alpha_x^2-\alpha_y^2)\epsilon_{yz}+2\alpha_x\alpha_y\epsilon_{xz}\right]+b_{34}\left[\frac{1}{2}\alpha_y\alpha_z(\epsilon_{xx}-\epsilon_{yy})+2\alpha_x\alpha_z\epsilon_{xy}\right].
\label{eq:E_me_trig_I}     
\end{aligned}
\end{equation}
Next, we obtain the equilibrium strains via Eq.\ref{eq:dE}. Replacing them into Eq.\ref{eq:delta_l} leads to \cite{Cullen}
\begin{equation}
\begin{aligned}
     \frac{\Delta l}{l_0}\Bigg\vert_{\boldsymbol{\beta}}^{\boldsymbol{\alpha}} & =\lambda^{\alpha1,0}(\beta_x^2+\beta_y^2)+\lambda^{\alpha2,0}\beta_z^2+\lambda^{\alpha1,2}\left(\alpha_z^2-\frac{1}{3}\right)(\beta_x^2+\beta_y^2)\\
     & + \lambda^{\alpha2,2}\left(\alpha_z^2-\frac{1}{3}\right)\beta_z^2+\lambda^{\gamma,1}\left[\frac{1}{2}(\alpha_x^2-\alpha_y^2)(\beta_x^2-\beta_y^2)+2\alpha_x\alpha_y\beta_x\beta_y\right]\\
     & + \lambda^{\gamma,2}(\alpha_x\alpha_z\beta_x\beta_z+\alpha_y\alpha_z\beta_y\beta_z)+\lambda_{12}\left[\frac{1}{2}\alpha_y\alpha_z(\beta_x^2-\beta_y^2)+\alpha_x\alpha_z\beta_x\beta_y\right]\\
     & +\lambda_{21}\left[\frac{1}{2}(\alpha_x^2-\alpha_y^2)\beta_y\beta_z+\alpha_x\alpha_y\beta_x\beta_z\right],
    \label{eq:delta_l_trig_I}
\end{aligned}
\end{equation}
where
\begin{equation}
    \begin{aligned}
        \lambda^{\alpha1,0} & = \frac{b_{11}C_{33}+b_{12}C_{13}}{C_{33}(C_{11}+C_{12})-2C_{13}^2},\\
        \lambda^{\alpha2,0} & = \frac{2b_{11}C_{13}-b_{12}(C_{11}+C_{12})}{C_{33}(C_{11}+C_{12})-2C_{13}^2},\\
        \lambda^{\alpha1,2} & = \frac{-b_{21}C_{33}+b_{22}C_{13}}{C_{33}(C_{11}+C_{12})-2C_{13}^2},\\
        \lambda^{\alpha2,2} & = \frac{2b_{21}C_{13}-b_{22}(C_{11}+C_{12})}{C_{33}(C_{11}+C_{12})-2C_{13}^2},\\
        \lambda^{\gamma,1} & = \frac{C_{14}b_{14}-C_{44}b_{3}}{\frac{1}{2}C_{44}(C_{11}-C_{12})-C_{14}^2},\\
        \lambda^{\gamma,2} & = \frac{\frac{1}{2}b_4(C_{11}-C_{12})-b_{34}C_{14}}{\frac{1}{2}C_{44}(C_{11}-C_{12})-C_{14}^2},\\
        \lambda_{12} & = \frac{C_{14}b_{4}-C_{44}b_{34}}{\frac{1}{2}C_{44}(C_{11}-C_{12})-C_{14}^2},\\
        \lambda_{21} & = \frac{\frac{1}{2}b_{14}(C_{11}-C_{12})-b_{3}C_{14}}{\frac{1}{2}C_{44}(C_{11}-C_{12})-C_{14}^2}.
    \label{eq:lamb_trig}
    \end{aligned}
\end{equation}
The MAE in an unstrained trigonal crystal up to fourth-order in $\alpha$ is the same to the hexagonal case (Eq.\ref{eq:E_mca_hex}).

\subsection{Tetragonal (I)}
\label{subsection:tetragonal_I}

\subsubsection{Single crystal}

The tetragonal (I) crystal system (point groups $4mm$, $422$, $\bar{4}2m$, $4/mmm$) has the following elastic stiffness tensor
\begin{equation}
\begin{aligned}
C^{tet(I)} =\begin{pmatrix}
C_{11} & C_{12} & C_{13} & 0 & 0 & 0 \\
C_{12} & C_{11} & C_{13}& 0 & 0 & 0 \\
C_{13} & C_{13} & C_{33} & 0& 0 & 0 \\
0 & 0 & 0 & C_{44} & 0 & 0 \\
0 & 0 & 0 & 0 & C_{44} & 0 \\
0 & 0 & 0 & 0 & 0 & C_{66} \\
\end{pmatrix},
\label{eq:stiffnees_tensor_tet}
\end{aligned}
\end{equation}
Hence, it has six independent elastic
constants $C_{11}$, $C_{12}$, $C_{13}$, $C_{33}$, $C_{44}$ and $C_{66}$. The elastic energy is given by
\begin{equation}
\begin{aligned}
 \frac{ E_{el}^{tet(I)}-E_0}{V_0} & = \frac{1}{2}C_{11}(\tilde{\epsilon}_{1}^2+\tilde{\epsilon}_{2}^2)+C_{12}\tilde{\epsilon}_{1}\tilde{\epsilon}_{2}+C_{13}(\tilde{\epsilon}_{1}+\tilde{\epsilon}_{2})\tilde{\epsilon}_{3}+\frac{1}{2}C_{33}\tilde{\epsilon}_{3}^2 \\
  & +\frac{1}{2}C_{44}(\tilde{\epsilon}_{4}^2+\tilde{\epsilon}_{5}^2)+\frac{1}{2}C_{66}\tilde{\epsilon}_{6}^2\\
 & = \frac{1}{2}c_{xxxx}(\epsilon_{xx}^2+\epsilon_{yy}^2)+c_{xxyy}\epsilon_{xx}\epsilon_{yy}+c_{xxzz}(\epsilon_{xx}+\epsilon_{yy})\epsilon_{zz}+\frac{1}{2}c_{zzzz}\epsilon_{zz}^2 \\
  & +2c_{yzyz}(\epsilon_{yz}^2+\epsilon_{xz}^2)+2c_{xyxy}\epsilon_{xy}^2
\end{aligned}
\label{eq:E_el_tet}
\end{equation}
where $C_{11}=c_{xxxx}$, $C_{12}=c_{xxyy}$, $C_{13}=c_{xxzz}$, $C_{33}=c_{zzzz}$, $C_{44}=c_{yzyz}$ and $C_{66}=c_{xyxy}$. On the other hand, there are 7 independent magnetoelastic constants \cite{Callen}. The magnetoelastic energy can be written as \cite{Frit,Cullen}
\begin{equation}
\begin{aligned}
    \frac{E_{me}^{tet(I)}}{V_0}  & =  b_{11}(\epsilon_{xx}+\epsilon_{yy})+b_{12}\epsilon_{zz}+b_{21}\left(\alpha_z^2-\frac{1}{3}\right)(\epsilon_{xx}+\epsilon_{yy})+b_{22}\left(\alpha_z^2-\frac{1}{3}\right)\epsilon_{zz}\\
    & + \frac{1}{2}b_3(\alpha_x^2-\alpha_y^2)(\epsilon_{xx}-\epsilon_{yy})+2b_3'\alpha_x\alpha_y\epsilon_{xy}+2b_4(\alpha_x\alpha_z\epsilon_{xz}+\alpha_y\alpha_z\epsilon_{yz}).
\label{eq:E_me_tet_I}     
\end{aligned}
\end{equation}
After the equilibrium strains are calculated by minimizing Eqs.\ref{eq:E_el_tet} and \ref{eq:E_me_tet_I} through Eq.\ref{eq:dE} and replaced into Eq.\ref{eq:delta_l}, we have \cite{Cullen}
\begin{equation}
\begin{aligned}
     \frac{\Delta l}{l_0}\Bigg\vert_{\boldsymbol{\beta}}^{\boldsymbol{\alpha}} & =\lambda^{\alpha1,0}(\beta_x^2+\beta_y^2)+\lambda^{\alpha2,0}\beta_z^2+\lambda^{\alpha1,2}\left(\alpha_z^2-\frac{1}{3}\right)(\beta_x^2+\beta_y^2)\\
     & + \lambda^{\alpha2,2}\left(\alpha_z^2-\frac{1}{3}\right)\beta_z^2+\frac{1}{2}\lambda^{\gamma,2}(\alpha_x^2-\alpha_y^2)(\beta_x^2-\beta_y^2)+2\lambda^{\delta,2}\alpha_x\alpha_y\beta_x\beta_y\\
     & + 2\lambda^{\epsilon,2}(\alpha_x\alpha_z\beta_x\beta_z+\alpha_y\alpha_z\beta_y\beta_z),
    \label{eq:delta_l_tet_I}
\end{aligned}
\end{equation}
where
\begin{equation}
    \begin{aligned}
        \lambda^{\alpha1,0} & = \frac{b_{11}C_{33}+b_{12}C_{13}}{C_{33}(C_{11}+C_{12})-2C_{13}^2},\\
        \lambda^{\alpha2,0} & = \frac{2b_{11}C_{13}-b_{12}(C_{11}+C_{12})}{C_{33}(C_{11}+C_{12})-2C_{13}^2},\\
        \lambda^{\alpha1,2} & = \frac{-b_{21}C_{33}+b_{22}C_{13}}{C_{33}(C_{11}+C_{12})-2C_{13}^2},\\
        \lambda^{\alpha2,2} & = \frac{2b_{21}C_{13}-b_{22}(C_{11}+C_{12})}{C_{33}(C_{11}+C_{12})-2C_{13}^2},\\
        \lambda^{\gamma,2} & = \frac{-b_3}{C_{11}-C_{12}},\\
        \lambda^{\delta,2} & = \frac{-b_3'}{2C_{66}},\\
        \lambda^{\epsilon,2} & = \frac{-b_4}{2C_{44}}.
    \label{eq:lamb_tet}
    \end{aligned}
\end{equation}
Mason derived an equivalent equation to Eq.\ref{eq:delta_l_tet_I} using a different arrangement of the terms and definitions of the magnetostrictive coefficients \cite{Mason}. The conversion formulas between the  magnetostrictive coefficients in Eq.\ref{eq:delta_l_tet_I} and those defined by Mason are shown in \ref{app_tet_I}.  The MAE in an unstrained tetragonal crystal up to fourth-order in $\alpha$ is the same to the hexagonal case (Eq.\ref{eq:E_mca_hex}).

\subsection{Orthorhombic}
\label{subsection:orthorhombic}

\subsubsection{Single crystal}

The orthorhombic crystal system (point groups $222$, $2mm$, $mmm$) has 9 independent elastic
constants $C_{11}$, $C_{12}$, $C_{13}$, $C_{22}$, $C_{23}$, $C_{33}$, $C_{44}$, $C_{55}$ and $C_{66}$, its elastic stiffness matrix reads\cite{mouhat,AELAS}
\begin{equation}
\begin{aligned}
C^{ortho} =\begin{pmatrix}
C_{11} & C_{12} & C_{13} & 0 & 0 & 0 \\
C_{12} & C_{22} & C_{23}& 0 & 0 & 0 \\
C_{13} & C_{23} & C_{33} & 0& 0 & 0 \\
0 & 0 & 0 & C_{44} & 0 & 0 \\
0 & 0 & 0 & 0 & C_{55} & 0 \\
0 & 0 & 0 & 0 & 0 & C_{66} \\
\end{pmatrix},
\label{eq:stiffnees_tensor_ort}
\end{aligned}
\end{equation}
Hence, inserting it into Eq.\ref{eq:E_el} leads to the following expression for the elastic energy 
\begin{equation}
\begin{aligned}
  \frac{E_{el}^{ortho}-E_{0}}{V_0} & = \frac{1}{2}C_{11}\tilde{\epsilon}_{1}^2+\frac{1}{2}C_{22}\tilde{\epsilon}_{2}^2+C_{12}\tilde{\epsilon}_{1}\tilde{\epsilon}_{2}+C_{13}\tilde{\epsilon}_{1}\tilde{\epsilon}_{3}+C_{23}\tilde{\epsilon}_{2}\tilde{\epsilon}_{3}+\frac{1}{2}C_{33}\tilde{\epsilon}_{3}^2 \\
  & +\frac{1}{2}C_{44}\tilde{\epsilon}_{4}^2+\frac{1}{2}C_{55}\tilde{\epsilon}_{5}^2+\frac{1}{2}C_{66}\tilde{\epsilon}_{6}^2 \\  
& = \frac{1}{2}c_{xxxx}\epsilon_{xx}^2+\frac{1}{2}c_{yyyy}\epsilon_{yy}^2+c_{xxyy}\epsilon_{xx}\epsilon_{yy}+c_{xxzz}\epsilon_{xx}\epsilon_{zz}+c_{yyzz}\epsilon_{yy}\epsilon_{zz} \\
  & +\frac{1}{2}c_{zzzz}\epsilon_{zz}^2+2c_{yzyz}\epsilon_{yz}^2+2c_{xzxz}\epsilon_{xz}^2+2c_{xyxy}\epsilon_{xy}^2.  
\end{aligned}
\label{eq:E_el_ortho}
\end{equation}
where $C_{11}=c_{xxxx}$, $C_{22}=c_{yyyy}$, $C_{12}=c_{xxyy}$, $C_{13}=c_{xxzz}$, $C_{23}=c_{yyzz}$, $C_{33}=c_{zzzz}$, $C_{44}=c_{yzyz}$, $C_{55}=c_{xzxz}$ and $C_{66}=c_{xyxy}$. The magnetoelastic energy contains 12 independent magnetoelastic constants \cite{Callen}. Mason derived the following expression of the relative length change \cite{Mason}
\begin{equation}
\begin{aligned}
     \frac{\Delta l}{l_0}\Bigg\vert_{\boldsymbol{\beta}}^{\boldsymbol{\alpha}} & =\lambda^{\alpha1,0}\beta_x^2+\lambda^{\alpha2,0}\beta_y^2+\lambda^{\alpha3,0}\beta_z^2+\lambda_1(\alpha_x^2\beta_x^2-\alpha_x\alpha_y\beta_x\beta_y-\alpha_x\alpha_z\beta_x\beta_z)\\
     & +\lambda_2(\alpha_y^2\beta_x^2-\alpha_x\alpha_y\beta_x\beta_y)+\lambda_3(\alpha_x^2\beta_y^2-\alpha_x\alpha_y\beta_x\beta_y)\\
     & +\lambda_4(\alpha_y^2\beta_y^2-\alpha_x\alpha_y\beta_x\beta_y-\alpha_y\alpha_z\beta_y\beta_z)+\lambda_5(\alpha_x^2\beta_z^2-\alpha_x\alpha_z\beta_x\beta_z)\\
     & + \lambda_6(\alpha_y^2\beta_z^2-\alpha_y\alpha_z\beta_y\beta_z)+4\lambda_7\alpha_x\alpha_y\beta_x\beta_y+4\lambda_8\alpha_x\alpha_z\beta_x\beta_z+4\lambda_9\alpha_y\alpha_z\beta_y\beta_z.
    \label{eq:delta_l_ortho}
\end{aligned}
\end{equation}
Note that we added the terms that describes the volume magnetostriction ($\lambda^{\alpha1,0}$, $\lambda^{\alpha2,0}$ and $\lambda^{\alpha3,0}$), which were not included in the original work of Mason \cite{Mason}. The expression of the magnetoelastic energy and the relations between magnetostrictive coefficients, elastic and magnetoelastic constants were not shown by Mason either. For completeness, here we deduce it from Eqs. \ref{eq:E_el_ortho} and \ref{eq:delta_l_ortho}. To do so, we aim to find the unknown functions $g_i$ and $f_i$ in the general form of the magnetoelastic energy in cartesian coordinates given by Eq. \ref{eq:E_me}. Firstly, we minimize Eqs. \ref{eq:E_el_ortho} and \ref{eq:E_me} via Eq. \ref{eq:dE}. This gives a set of equations that links the unknown functions $g_i$ and $f_i$ with the equilibrium strains. Next, we extract the equilibrium strains by direct comparison between Eqs. \ref{eq:delta_l} and \ref{eq:delta_l_ortho}. Finally, we substitute the equilibrium strains into the set of equations that relates $g_i$ and $f_i$ with the equilibrium strains, from which we obtain $g_i$ and $f_i$. Inserting the calculated $g_i$ and $f_i$ into Eq. \ref{eq:E_me} we have
\begin{equation}
\begin{aligned}
    \frac{E_{me}^{ortho}}{V_0}  & =  b_{01}\epsilon_{xx}+b_{02}\epsilon_{yy}+b_{03}\epsilon_{zz}+b_1\alpha_x^2\epsilon_{xx}+b_2\alpha_y^2\epsilon_{xx}+b_3\alpha_x^2\epsilon_{yy}+b_4\alpha_y^2\epsilon_{yy}\\
    & +b_5\alpha_x^2\epsilon_{zz}+b_6\alpha_y^2\epsilon_{zz}+2b_7\alpha_x\alpha_y\epsilon_{xy}+2b_8\alpha_x\alpha_z\epsilon_{xz}+2b_9\alpha_y\alpha_z\epsilon_{yz},
\label{eq:E_me_ortho}     
\end{aligned}
\end{equation}
where
\begin{equation}
    \begin{aligned}
        b_{01} & = -C_{11} \lambda^{\alpha1,0} - C_{12} \lambda^{\alpha2,0} - C_{13} \lambda^{\alpha3,0} \\
        b_{02} & = -C_{12} \lambda^{\alpha1,0} - C_{22} \lambda^{\alpha2,0} - C_{23} \lambda^{\alpha3,0}\\
        b_{03} & = -C_{13} \lambda^{\alpha1,0} - C_{23} \lambda^{\alpha2,0} - C_{33} \lambda^{\alpha3,0} \\
        b_1 & = -C_{11} \lambda_1 - C_{12} \lambda_3 - C_{13} \lambda_5 \\
        b_2 & = -C_{11} \lambda_2 - C_{12} \lambda_4 - C_{13} \lambda_6 \\
        b_3 & = -C_{12} \lambda_1 - C_{22} \lambda_3 - C_{23} \lambda_5 \\
        b_4 & = -C_{12} \lambda_2 - C_{22} \lambda_4 - C_{23} \lambda_6 \\
        b_5 & = -C_{13} \lambda_1 - C_{23} \lambda_3 - C_{33} \lambda_5 \\
        b_6 & = -C_{13} \lambda_2 - C_{23} \lambda_4 - C_{33} \lambda_6 \\
        b_7 & = C_{66} (\lambda_1 + \lambda_2 + \lambda_3 + \lambda_4 - 4\lambda_7) \\
        b_8 & = C_{55} (\lambda_1 + \lambda_5 - 4\lambda_8) \\
        b_9 & = C_{44} (\lambda_4 + \lambda_6 - 4\lambda_9).
    \label{eq:lamb_ortho}
    \end{aligned}
\end{equation}
Alternatively, one can deduce the magnetoelastic energy using the general approach based on the symmetry strains and direction cosine polynomial for each irreducible representation \cite{Callen,CLARK1980531,Cullen}. This approach may lead to different definitions of the magnetoelastic constants and magnetostrictive coefficients, as we have discussed for the hexagonal (I) and tetragonal (I) systems in \ref{app:hex_I} and \ref{app_tet_I}, respectively. A generalization of the approach taken by Becker and Doring \cite{Becker} for orthorhombic crystals can be found in Ref. \cite{Carr}. The MAE in an unstrained orthorhombic crystal up to fourth-order in $\alpha$ is\cite{Mason}
\begin{equation}
\begin{aligned}
    \frac{E_{K}^0}{V_0}= K_0+K_1\alpha_x^2+K_2\alpha_y^2. 
\label{eq:E_mca_ortho}     
\end{aligned}
\end{equation}

\section{Methodology}
\label{section:methodology}

\subsection{Calculation of magnetostrictive coefficients and magnetoelastic constants}
\label{subsection:methodology_coeff}

The methodology implemented in the program MAELAS to calculate the anisotropic magnetostrictive coefficients is a generalization of the approach proposed by Wu and Freeman for cubic crystals \cite{Wu1996,WU1997}. In this method,  one  measuring length direction  $\boldsymbol{\beta}^i$ and two magnetization directions ($\boldsymbol{\alpha}_1^i$ and $\boldsymbol{\alpha}_2^i$) are chosen for each magnetostrictive coefficient ($\lambda^i$) in such a way that
\begin{equation}
     \frac{\Delta l}{l_0}\Bigg\vert_{\boldsymbol{\beta}^i}^{\boldsymbol{\alpha}_1^i} -  \frac{\Delta l}{l_0}\Bigg\vert_{\boldsymbol{\beta}^i}^{\boldsymbol{\alpha}_2^i} = \rho^i \lambda^i,
    \label{eq:delta_l_method}
\end{equation}
where $\rho^i$ is a real number. In Table \ref{tab:beta_alpha_data} we show the selected set of  $\boldsymbol{\beta}^i$, $\boldsymbol{\alpha}_1^i$ and $\boldsymbol{\alpha}_2^i$ in MAELAS that fulfils Eq.\ref{eq:delta_l_method} for each $\rho^i$. Next, the left hand side of Eq.\ref{eq:delta_l_method} is written as
\begin{equation}
\begin{aligned}
     \frac{\Delta l}{l_0}\Bigg\vert_{\boldsymbol{\beta}^i}^{\boldsymbol{\alpha}_1^i} -  \frac{\Delta l}{l_0}\Bigg\vert_{\boldsymbol{\beta}^i}^{\boldsymbol{\alpha}_2^i} & = \frac{ l_1 -l_0}{l_0} -  \frac{l_2-l_0}{l_0} = \frac{2 (l_1 -l_2)}{(l_1+l_2)\left[1-\frac{l_1+l_2-2l_0}{l_1+l_2}\right]}\\
     & = \frac{2 (l_1 -l_2)}{l_1+l_2}\left[1+\frac{l_1+l_2-2l_0}{l_1+l_2}+...\right]\approx \frac{2 (l_1 -l_2)}{l_1+l_2},
    \label{eq:delta_l_method_approx}
    \end{aligned}
\end{equation}
where in the last approximation we assume $\vert l_{1(2)}-l_0\vert/l_0\ll 1$. This assumption is reasonable for all known magnetostrictive materials. For instance, a very large value of $\Delta l/l_0$ is about $4.5\times10^{-3}$ found in TbFe$_2$ (Laves phase C15) along direction [111] at $T=0$K \cite{Eng}, where this approximation is fine. This approximation allows to get rid of $l_0$ (length along $\boldsymbol{\beta}$ in the macroscopic demagnetized state) which can't be calculated with DFT methods easily. Combining Eqs.\ref{eq:delta_l_method} and \ref{eq:delta_l_method_approx} one can write the magnetostrictive coefficients as
\begin{equation}
     \lambda^i=\frac{2 (l_1 -l_2)}{\rho^i (l_1+l_2)},
    \label{eq:lambda_method}
\end{equation}
where the value of $\rho^i$ for each $\lambda^i$ is given in Table \ref{tab:beta_alpha_data}. The quantities $l_1$ and $l_2$ correspond to the cell length along $\boldsymbol{\beta}$ when the magnetization points to $\boldsymbol{\alpha}_1$ and $\boldsymbol{\alpha}_2$, respectively, and  are calculated through an optimization of the energy. Namely, a set of deformed unit cells is firstly generated using the deformation modes described in  \ref{app_matrix}.  For each deformed cell, the energy is calculated constraining the spins to the directions given by $\boldsymbol{\alpha}_1$ and $\boldsymbol{\alpha}_2$. Next, the energy versus the cell length along $\boldsymbol{\beta}$ for each spin direction $\boldsymbol{\alpha}_{1(2)}$ is fitted to a quadratic polynomial
\begin{equation}
      E(\boldsymbol{\alpha}_j,l)=A_j l^2+B_j l+C_j,\quad j=1,2
    \label{eq:E_fit}
\end{equation}
where $A_j$, $B_j$ and $C_j$ ($j=1,2$) are fitting parameters. The minimum of this function for spin direction $\boldsymbol{\alpha}_{1(2)}$ corresponds to  $l_{1(2)}=-B_{1(2)}/(2A_{1(2)})$. Once $l_1$ and $l_2$ are determined, one obtains the magnetostrictive coefficients using Eq.\ref{eq:lambda_method}. The magnetostrictive coefficients can also be written in terms of the derivative of the energy with respect $l$ evaluated at $l=l_2$ as \cite{Wu}
\begin{equation}
     \lambda^i\approx-\frac{1}{\eta^i B_1}\cdot\frac{\partial[ E(\boldsymbol{\alpha}_2,l)-E(\boldsymbol{\alpha}_1,l)]}{\partial l}\Bigg\vert_{l=l_2}
    \label{eq:lambda_derivative}
\end{equation}
where $B_1$ is always negative.  In Table \ref{tab:beta_alpha_data}, we see that our choice of $\boldsymbol{\beta}$ and $\boldsymbol{\alpha}_{1(2)}$ makes   $\rho$ depend on some magnetostrictive coefficients for $\lambda_7$, $\lambda_8$ and $\lambda_9$ in orthorhombic crystals. For instance, working out the coefficient $\lambda_7$ via Eq.\ref{eq:lambda_method} we have
\begin{equation}
     \lambda_7=\frac{(a^2+b^2) (l_1 -l_2)}{ab(l_1+l_2)}-\frac{(a-b)(a[\lambda_1+\lambda_2]-b[\lambda_3+\lambda_4])}{4ab},
    \label{eq:lambda_7}
\end{equation}
where $a$ and $b$ are the relaxed (not distorted) lattice parameters of the unit cell. Here, MAELAS makes use of the values of  $\lambda_1$, $\lambda_2$, $\lambda_3$ and $\lambda_4$ calculated previously in order to compute $\lambda_7$. Note that a simpler expression for $\lambda_7$ can be achieved choosing the measuring length direction $\boldsymbol{\beta}=\left(\frac{1}{\sqrt{2}},\frac{1}{\sqrt{2}},0\right)$. However, from a computational point of view, it is easy to extract the cell length $l$ along $\boldsymbol{\beta}=\left(\frac{a}{\sqrt{a^2+b^2}},\frac{b}{\sqrt{a^2+b^2}},0\right)$ of each deformed cell generated with the deformation gradients discussed in \ref{app_matrix}. Similarly, one can deduce the explicit equation for $\lambda_8$ and $\lambda_9$.

Lastly, if the elastic tensor is provided in the format given by the program AELAS \cite{AELAS}, then the magnetoelastic constants ($b_k$) are also calculated from the relations $b_k=b_k(\lambda^i,C_{nm})$ given in Section \ref{section:theory}.

\begin{table}[]
\centering
\caption{Selected cell length ($\boldsymbol{\beta}$) and magnetization directions ($\boldsymbol{\alpha}_1$, $\boldsymbol{\alpha}_2$) in MAELAS to calculate the anisotropic magnetostrictive coefficients according to Eq.\ref{eq:delta_l_method}. The first column shows the crystal system and the corresponding lattice convention set in MAELAS based on the IEEE format \cite{AELAS}. The second column presents the equation of the relative length change that we used in Eq.\ref{eq:delta_l_method} for each crystal system. In the last column we show the values of the parameter $\rho$ that is defined in Eq.\ref{eq:delta_l_method}. The symbols $a,b,c$ correspond to the lattice parameters of the relaxed (not distorted) unit cell.}
\label{tab:beta_alpha_data}
\resizebox{\textwidth}{!}{%
\begin{tabular}{@{}lllllll@{}}
\toprule
Crystal system & \multicolumn{1}{c}{$\frac{\Delta l}{l_0}$} &
  \multicolumn{1}{c}{\begin{tabular}[c]{@{}c@{}}Magnetostrictive \\ coefficient\end{tabular}} &
  \multicolumn{1}{c}{$\boldsymbol{\beta}$} &
  \multicolumn{1}{c}{$\boldsymbol{\alpha}_1$} &
  \multicolumn{1}{c}{$\boldsymbol{\alpha}_2$} &
  \multicolumn{1}{c}{$\rho$} \\ \hline\hline
Cubic (I)     & Eq.\ref{eq:delta_l_cub_I} & \multicolumn{1}{c}{$\lambda_{001}$} &  \multicolumn{1}{c}{$(0,0,1)$}& \multicolumn{1}{c}{$(0,0,1)$} & \multicolumn{1}{c}{$(1,0,0)$} & \multicolumn{1}{c}{$\frac{3}{2}$}\\
  $\boldsymbol{a}\|\hat{\boldsymbol{x}}$, $\boldsymbol{b}\|\hat{\boldsymbol{y}}$, $\boldsymbol{c}\|\hat{\boldsymbol{z}}$          &  & \multicolumn{1}{c}{$\lambda_{111}$} & \multicolumn{1}{c}{$\left(\frac{1}{\sqrt{3}},\frac{1}{\sqrt{3}},\frac{1}{\sqrt{3}}\right)$} & \multicolumn{1}{c}{$\left(\frac{1}{\sqrt{3}},\frac{1}{\sqrt{3}},\frac{1}{\sqrt{3}}\right)$} & \multicolumn{1}{c}{$\left(\frac{1}{\sqrt{2}},0,\frac{-1}{\sqrt{2}}\right)$} &  \multicolumn{1}{c}{$\frac{3}{2}$}\\ \hline
Hexagonal (I) & Eq.\ref{eq:delta_l_hex_I}  & \multicolumn{1}{c}{$\lambda^{\alpha1,2}$}    & \multicolumn{1}{c}{$\left(1,0,0\right)$}                              & \multicolumn{1}{c}{$\left(\frac{1}{\sqrt{3}},\frac{1}{\sqrt{3}},\frac{1}{\sqrt{3}}\right)$} & \multicolumn{1}{c}{$\left(\frac{1}{\sqrt{2}},\frac{1}{\sqrt{2}},0\right)$} & \multicolumn{1}{c}{$\frac{1}{3}$}   \\
   $\boldsymbol{a}\|\hat{\boldsymbol{x}}$,  $\boldsymbol{c}\|\hat{\boldsymbol{z}}$        &     &    \multicolumn{1}{c}{$\lambda^{\alpha2,2}$}                                  & \multicolumn{1}{c}{$(0,0,1)$} & \multicolumn{1}{c}{$(0,0,1)$} & \multicolumn{1}{c}{$(1,0,0)$} & \multicolumn{1}{c}{$1$} \\
    $\boldsymbol{b}=\left(-\frac{a}{2},\frac{\sqrt{3}a}{2},0\right)$        &     &    \multicolumn{1}{c}{$\lambda^{\gamma,2}$}                                  & \multicolumn{1}{c}{$(1,0,0)$} & \multicolumn{1}{c}{$(1,0,0)$} & \multicolumn{1}{c}{$(0,1,0)$} & \multicolumn{1}{c}{$1$}  \\
    $a=b\neq c$      &     & \multicolumn{1}{c}{$\lambda^{\epsilon,2}$}    & \multicolumn{1}{c}{$\frac{\left(a,0,c\right)}{\sqrt{a^2+c^2}}$}                             & \multicolumn{1}{c}{$\left(\frac{1}{\sqrt{2}},0,\frac{1}{\sqrt{2}}\right)$} & \multicolumn{1}{c}{$\left(\frac{-1}{\sqrt{2}},0,\frac{1}{\sqrt{2}}\right)$} & \multicolumn{1}{c}{$\frac{2ac}{a^2+c^2}$} \\ \hline
Trigonal (I)  & Eq.\ref{eq:delta_l_trig_I} & \multicolumn{1}{c}{$\lambda^{\alpha1,2}$}    & \multicolumn{1}{c}{$\left(1,0,0\right)$}                              & \multicolumn{1}{c}{$\left(0,0,1\right)$} & \multicolumn{1}{c}{$\left(\frac{1}{\sqrt{2}},\frac{1}{\sqrt{2}},0\right)$} & \multicolumn{1}{c}{$1$}   \\
      $\boldsymbol{a}\|\hat{\boldsymbol{x}}$,  $\boldsymbol{c}\|\hat{\boldsymbol{z}}$      &    &    \multicolumn{1}{c}{$\lambda^{\alpha2,2}$}                                  & \multicolumn{1}{c}{$(0,0,1)$} & \multicolumn{1}{c}{$(0,0,1)$} & \multicolumn{1}{c}{$(1,0,0)$} & \multicolumn{1}{c}{$1$} \\
 $\boldsymbol{b}=\left(-\frac{a}{2},\frac{\sqrt{3}a}{2},0\right)$           &    &    \multicolumn{1}{c}{$\lambda^{\gamma,1}$}                                  & \multicolumn{1}{c}{$(1,0,0)$} & \multicolumn{1}{c}{$(1,0,0)$} & \multicolumn{1}{c}{$(0,1,0)$} & \multicolumn{1}{c}{$1$}  \\
     $a=b\neq c$      &    & \multicolumn{1}{c}{$\lambda^{\gamma,2}$}    & \multicolumn{1}{c}{$\frac{\left(a,0,c\right)}{\sqrt{a^2+c^2}}$}                             & \multicolumn{1}{c}{$\left(\frac{1}{\sqrt{2}},0,\frac{1}{\sqrt{2}}\right)$} & \multicolumn{1}{c}{$\left(\frac{1}{\sqrt{2}},0,\frac{-1}{\sqrt{2}}\right)$} & \multicolumn{1}{c}{$\frac{ac}{a^2+c^2}$} \\
           &     & \multicolumn{1}{c}{$\lambda_{12}$}    & \multicolumn{1}{c}{$\frac{\left(a,0,c\right)}{\sqrt{a^2+c^2}}$}                             & \multicolumn{1}{c}{$\left(0,\frac{1}{\sqrt{2}},\frac{1}{\sqrt{2}}\right)$} & \multicolumn{1}{c}{$\left(0,\frac{1}{\sqrt{2}},\frac{-1}{\sqrt{2}}\right)$} & \multicolumn{1}{c}{$\frac{a^2}{2(a^2+c^2)}$} \\
            &    & \multicolumn{1}{c}{$\lambda_{21}$}    &  \multicolumn{1}{c}{$\frac{\left(a,0,c\right)}{\sqrt{a^2+c^2}}$}                            & \multicolumn{1}{c}{$\left(\frac{1}{\sqrt{2}},\frac{1}{\sqrt{2}},0\right)$} & \multicolumn{1}{c}{$\left(\frac{1}{\sqrt{2}},\frac{-1}{\sqrt{2}},0\right)$} & \multicolumn{1}{c}{$\frac{ac}{a^2+c^2}$}   \\ \hline
Tetragonal (I) & Eq.\ref{eq:delta_l_tet_I} & \multicolumn{1}{c}{$\lambda^{\alpha1,2}$}    & \multicolumn{1}{c}{$\left(1,0,0\right)$}                              & \multicolumn{1}{c}{$\left(\frac{1}{\sqrt{3}},\frac{1}{\sqrt{3}},\frac{1}{\sqrt{3}}\right)$} & \multicolumn{1}{c}{$\left(\frac{1}{\sqrt{2}},\frac{1}{\sqrt{2}},0\right)$} & \multicolumn{1}{c}{$\frac{1}{3}$}   \\
    $\boldsymbol{a}\|\hat{\boldsymbol{x}}$, $\boldsymbol{b}\|\hat{\boldsymbol{y}}$, $\boldsymbol{c}\|\hat{\boldsymbol{z}}$        &    &    \multicolumn{1}{c}{$\lambda^{\alpha2,2}$}                                  & \multicolumn{1}{c}{$(0,0,1)$} & \multicolumn{1}{c}{$(0,0,1)$} & \multicolumn{1}{c}{$(1,0,0)$} & \multicolumn{1}{c}{$1$} \\
     $a=b\neq c$      &    &    \multicolumn{1}{c}{$\lambda^{\gamma,2}$}                                  & \multicolumn{1}{c}{$(1,0,0)$} & \multicolumn{1}{c}{$(1,0,0)$} & \multicolumn{1}{c}{$(0,1,0)$} & \multicolumn{1}{c}{$1$}  \\
           &    & \multicolumn{1}{c}{$\lambda^{\epsilon,2}$}    & \multicolumn{1}{c}{$\frac{\left(a,0,c\right)}{\sqrt{a^2+c^2}}$}                             & \multicolumn{1}{c}{$\left(\frac{1}{\sqrt{2}},0,\frac{1}{\sqrt{2}}\right)$} & \multicolumn{1}{c}{$\left(\frac{-1}{\sqrt{2}},0,\frac{1}{\sqrt{2}}\right)$} & \multicolumn{1}{c}{$\frac{2ac}{a^2+c^2}$}  \\
         &       & \multicolumn{1}{c}{$\lambda^{\delta,2}$}    & \multicolumn{1}{c}{$\left(\frac{1}{\sqrt{2}},\frac{1}{\sqrt{2}},0\right)$}                             & \multicolumn{1}{c}{$\left(\frac{1}{\sqrt{2}},\frac{1}{\sqrt{2}},0\right)$} & \multicolumn{1}{c}{$\left(\frac{-1}{\sqrt{2}},\frac{1}{\sqrt{2}},0\right)$} & \multicolumn{1}{c}{$1$} \\ \hline
Orthorhombic & Eq.\ref{eq:delta_l_ortho} & \multicolumn{1}{c}{$\lambda_{1}$} &  \multicolumn{1}{c}{$(1,0,0)$}& \multicolumn{1}{c}{$(1,0,0)$} & \multicolumn{1}{c}{$(0,0,1)$} &   \multicolumn{1}{c}{$1$}  \\
  $\boldsymbol{a}\|\hat{\boldsymbol{x}}$, $\boldsymbol{b}\|\hat{\boldsymbol{y}}$, $\boldsymbol{c}\|\hat{\boldsymbol{z}}$          &    & \multicolumn{1}{c}{$\lambda_{2}$} &  \multicolumn{1}{c}{$(1,0,0)$}& \multicolumn{1}{c}{$(0,1,0)$} & \multicolumn{1}{c}{$(0,0,1)$} &   \multicolumn{1}{c}{$1$} \\
    $c<a<b$        &   & \multicolumn{1}{c}{$\lambda_{3}$} &  \multicolumn{1}{c}{$(0,1,0)$}& \multicolumn{1}{c}{$(1,0,0)$} & \multicolumn{1}{c}{$(0,0,1)$} &   \multicolumn{1}{c}{$1$} \\
             &  & \multicolumn{1}{c}{$\lambda_{4}$} &  \multicolumn{1}{c}{$(0,1,0)$}& \multicolumn{1}{c}{$(0,1,0)$} & \multicolumn{1}{c}{$(0,0,1)$} &   \multicolumn{1}{c}{$1$}  \\
             &  & \multicolumn{1}{c}{$\lambda_{5}$} &  \multicolumn{1}{c}{$(0,0,1)$}& \multicolumn{1}{c}{$(1,0,0)$} & \multicolumn{1}{c}{$(0,0,1)$} &   \multicolumn{1}{c}{$1$}  \\
             &  & \multicolumn{1}{c}{$\lambda_{6}$} &  \multicolumn{1}{c}{$(0,0,1)$}& \multicolumn{1}{c}{$(0,1,0)$} & \multicolumn{1}{c}{$(0,0,1)$} &   \multicolumn{1}{c}{$1$}  \\
              &  & \multicolumn{1}{c}{$\lambda_{7}$} &  \multicolumn{1}{c}{$\frac{\left(a,b,0\right)}{\sqrt{a^2+b^2}}$}    & \multicolumn{1}{c}{$\left(\frac{1}{\sqrt{2}},\frac{1}{\sqrt{2}},0\right)$}  & \multicolumn{1}{c}{$(0,0,1)$} &  \multicolumn{1}{c}{$\frac{(a-b)(a[\lambda_1 + \lambda_2] - b[\lambda_3 + \lambda_4]) + 4 a b \lambda_7}{2(a^2+b^2)\lambda_7}$}  \\
              & & \multicolumn{1}{c}{$\lambda_{8}$} &  \multicolumn{1}{c}{$\frac{\left(a,0,c\right)}{\sqrt{a^2+c^2}}$}    & \multicolumn{1}{c}{$\left(\frac{1}{\sqrt{2}},0,\frac{1}{\sqrt{2}}\right)$}  & \multicolumn{1}{c}{$(0,0,1)$} &  \multicolumn{1}{c}{$\frac{(a - c) (a \lambda_1 - c \lambda_5) + 4 a c \lambda_8}{2(a^2+c^2)\lambda_8}$}  \\
              & & \multicolumn{1}{c}{$\lambda_{9}$} &  \multicolumn{1}{c}{$\frac{\left(0,b,c\right)}{\sqrt{b^2+c^2}}$}    & \multicolumn{1}{c}{$\left(0,\frac{1}{\sqrt{2}},\frac{1}{\sqrt{2}}\right)$}  & \multicolumn{1}{c}{$(0,0,1)$} &  \multicolumn{1}{c}{$\frac{(b - c) (b \lambda_4 - c \lambda_6) + 4 b c \lambda_9 }{2(b^2+c^2)\lambda_9}$} \\ \bottomrule
\end{tabular}
}
\end{table}

\subsection{Program workflow}
\label{subsection:method_workflow}

\begin{figure}[ht!]
\centering
\includegraphics[width=1.0\columnwidth ,angle=0]{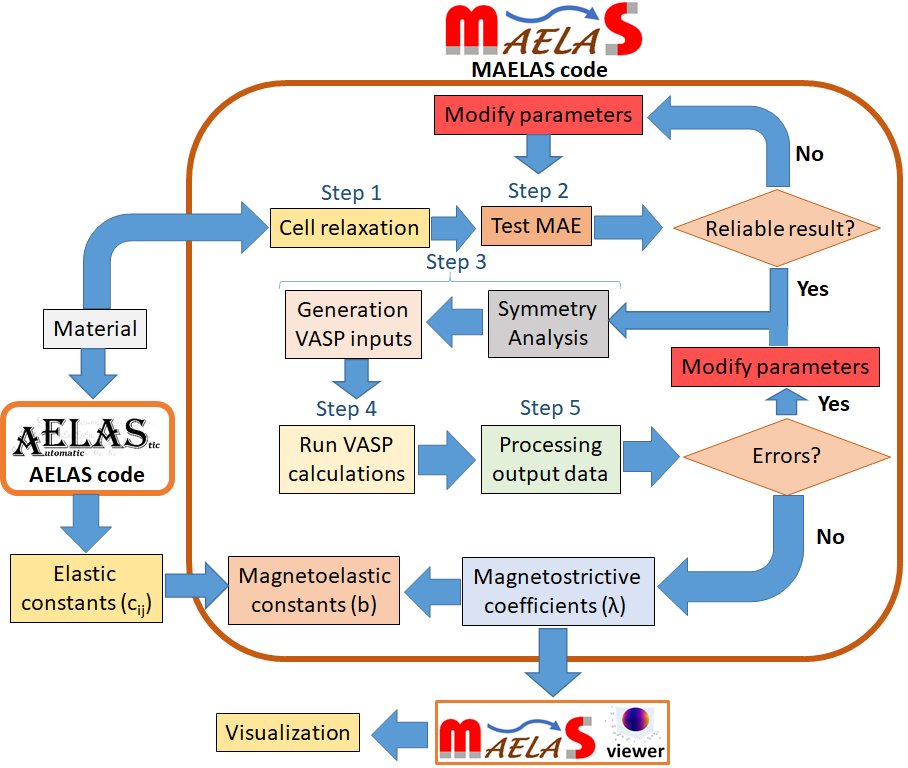}
\caption{Workflow of the program MAELAS and its connection with the program AELAS \cite{AELAS}.}
\label{fig:workflow}
\end{figure}

The program MAELAS has been designed to read and write files for the Vienna Ab initio Simulation Package (VASP) code \cite{vasp_1,vasp_2,vasp_3}. The workflow of MAELAS can be splitted into 5 steps: (i) cell relaxation, (ii) test of MAE, (iii) generation of distorted cells and spin directions, (iv) calculation of the energy with VASP, and (v) calculation of magnetostrictive coefficients and magnetoelastic constants. In Fig.\ref{fig:workflow} we show a diagram with a summary of the MAELAS workflow. In the first step, it performs a full cell relaxation (ionic positions, cell volume, and cell shape) of the input unit cell. If one wants to use non-relaxed lattice parameters (like experimental ones), then this step can be skipped. In the next step, it is recommended to check if it is possible to obtain a realistic value of MAE for the ground state (not distorted cell). To do so, MAELAS generates the VASP input files to calculate MAE for two spin directions given by the user which should be compared with available experimental data. In the third step, MAELAS performs a symmetry analysis with pymatgen library \cite{pymatgen} to determine the crystal system, redefine the structure using the same IEEE lattice convention as in AELAS \cite{AELAS}, and lastly generates a set of volume-conserving deformed unit cells and spin directions according to Table \ref{tab:beta_alpha_data}. Alternatively, the crystal symmetry of the system can also be imposed by the user manually, which could be useful in some cases. In the fourth step, one should run the VASP calculations using these inputs. In order to help in this task, MAELAS generates some bash scripts to run all calculations with VASP automatically. Lastly, MAELAS analyzes the calculated energies and fits them to a quadratic function (see Section \ref{subsection:methodology_coeff}) in order to obtain the magnetostrictive coefficients. If the elastic tensor is provided in the format given by the program AELAS \cite{AELAS}, then the magnetoelastic constants ($b_k$) are also calculated from the relations $b_k=b_k(\lambda^i,C_{nm})$ given in Section \ref{section:theory}. The obtained magnetostrictive coefficients and MAE can be further analyzed through the  online visualization tool called MAELASviewer that we have also developed \cite{maelasviewer,maelasviewerWeb,maelasviewerGithub}.

Detecting possible calculation failures on fly is a very important feature of an automated high-throughput code. For instance, MAELAS prints a warning message when the R-squared of the quadratic curve fitting is lower than 0.98. It also automatically generates figures showing the quadratic curve fitting and the energy difference between states with spin directions $\boldsymbol{\alpha}_2$ and $\boldsymbol{\alpha}_1$ versus the cell length along $\boldsymbol{\beta}$, so that the users can check the results easily.

\subsection{Computational details}
\label{subsection:comp_details}

The MAELAS code is written in Python3, and its source and documentation files are available in GitHub repository \cite{Maelas}. The DFT calculations are performed with VASP code, which is an implementation of the projector augmented wave (PAW) method \cite{vasp_4}. We use the interaction potentials generated for the Perdew-Burke-Ernzerhof (PBE) version \cite{Perdew} of the Generalized Gradient Approximation (GGA). We follow the recommended procedure to determine MAE with SOC included non-self-consistently. Namely, a first collinear spin-polarized job (without SOC) is performed to calculate the wavefunction and charge density, and then a second non-collinear spin-polarized job is performed in a non-self-consistent manner, by switching on the SOC, reading the wavefunction and charge density generated in the collinear job, and defining the spin orientation through the quantisation axis (SAXIS-tag) \cite{VASP_NCL}. The number of bands included in the non-collinear job is set twice as large as the number of bands in the collinear job. By default, the code sets the tetrahedron method with Bl\"ochl corrections for smearing in the calculation of MAE. The energy convergence criterion of the electronic self-consistency was chosen as $10^{-9}$  eV/cell, while the force convergence criterion of ionic relaxation was used, with all forces acting on atoms being lower than $10^{-3}$ eV/\r{A}. MAELAS also generates the input file with the set of k-points in the reciprocal space for VASP by using an automatic Monkhorst–Pack k-mesh \cite{Monk} gamma-centered grid with length parameter $R_k$ given by the user in the command line. Note the default settings generated by MAELAS for VASP might not work well for some materials, so that the user should check and tune them accordingly. For instance, in Section \ref{section:tests} we show some specific VASP settings and tests for few known materials. It is possible to use MAELAS with other codes instead of VASP, after file conversion to VASP format files. Although, this process might require some extra work for the user. For instance, we have recently made an interface between MAELAS and LAMMPS (spin-lattice simulations) \cite{nieves2020spinlattice,TRANCHIDA2018406,PLIMPTON19951}.

\section{Examples}
\label{section:tests}

In this section, we present some examples of the calculation of MAE, anisotropic magnetostrictive coefficients, elastic and magnetoelastic constants using MAELAS combined with AELAS \cite{AELAS} for a set of well-known magnetic materials. AELAS determines second-order elastic constants from the quadratic coefficients of
the polynomial fitting of the energies versus strain relationships efficiently. For each material, we split the analysis into two parts. Firstly, we perform a cell relaxation, evaluate MAE and compute magnetostriction with MAELAS. In the second stage, we calculate the elastic constants with AELAS, and we use them as inputs to compute the magnetoelastic constants with MAELAS. To do so, we follow the workflow  discussed in Section \ref{subsection:method_workflow} (see Fig. \ref{fig:workflow}). A summary of the results obtained in the following tests is shown in Tables \ref{tab:tests_MAELAS}, \ref{tab:tests_AELAS_1} and \ref{tab:tests_AELAS_2}. All calculations correspond to zero-temperature.

\begin{table}[H]
\centering
\caption{Anisotropic magnetostrictive coefficients and MAE calculated using the program MAELAS and measured in experiment ($T\approx0$ K) for a set of magnetic materials. In parenthesis we show the magnetostrictive coefficients with Mason's definitions obtained using the relations given by Eq. \ref{eq:lamb_Mason_hex_I}. These data correspond to the simulations with the largest number of k-points discussed in the main text.}
\label{tab:tests_MAELAS}
\resizebox{\textwidth}{!}{%
\begin{tabular}{@{}ccc|ccc|ccc@{}}
\toprule
Material &
  Crystal system &
  Method &
  \begin{tabular}[c]{@{}c@{}}Magnetostriction \\ coefficient\end{tabular} &
  \begin{tabular}[c]{@{}c@{}}MAELAS\\ ($\times10^{-6}$)\end{tabular} &
  \begin{tabular}[c]{@{}c@{}}Expt.\\ ($\times10^{-6}$)\end{tabular} &
  MAE &
  \begin{tabular}[c]{@{}c@{}}MAELAS\\ ($\mu$eV/atom)\end{tabular} &
  \begin{tabular}[c]{@{}c@{}}Expt.\\ ($\mu$eV/atom)\end{tabular} \\ \midrule \hline
FCC Ni      & Cubic (I)      & DFT GGA & $\lambda_{001}$        & -78.4 & -60$^a$   & $E(110)-E(001)$ & 0.03 & -2.15$^b$ \\
            &       SG 225         &     & $\lambda_{111}$        & -46.1 & -35$^a$  & $E(111)-E(001)$ & 0.34 &  -2.73$^b$  \\ & & & & & & & & \\
                &      & SD-MD & $\lambda_{001}$        & -61.9$^h$ &  & $E(110)-E(001)$ & -2.14$^h$ &  \\
            &               &     & $\lambda_{111}$        & -35.4$^h$ &   & $E(111)-E(001)$ & -2.86$^h$ &    \\ \midrule
BCC Fe      & Cubic (I)      & DFT GGA & $\lambda_{001}$        & 25.7 & 26$^a$   & $E(110)-E(001)$ & 0.24 & 1.0$^b$ \\
            &       SG 229         &     & $\lambda_{111}$        & 17.2 & -30$^a$  & $E(111)-E(001)$ & 0.32 &  1.34$^b$  \\  & & & & & & & & \\
     &      & SD-MD & $\lambda_{001}$        & 25.9$^h$ &   & $E(110)-E(001)$ & 0.99$^h$ & \\
            &              &     & $\lambda_{111}$        & -30.3$^h$ &   & $E(111)-E(001)$ & 1.33$^h$ &    \\ \midrule
HCP Co      & Hexagonal (I)  & DFT SCAN & $\lambda^{\alpha1,2}$ ($\lambda_A$) & 85 (-78) & 95 (-66)$^c$  & $E(100)-E(001)$ & 53 & 61$^b$\\
            &          SG 194      &     & $\lambda^{\alpha2,2}$ ($\lambda_B$) & -115 (-92) &  -126 (-123)$^c$ &                 &  &  \\
            &                &     & $\lambda^{\gamma,2}$  ($\lambda_C$) & 15 (115) & 57 (126)$^c$ &                 &  &  \\
            &                &     & $\lambda^{\epsilon,2}$ ($\lambda_D$) & -19 (-1) & -286 (-128)$^c$ &                 &  &  \\& & & & & & & & \\
               &  & DFT LSDA+U & $\lambda^{\alpha1,2}$ ($\lambda_A$) & 111 (-109) &   & $E(100)-E(001)$ & 58 & \\
            &               &  $J=0.8$eV & $\lambda^{\alpha2,2}$ ($\lambda_B$) & -251 (-114) &  &                 &  &  \\
            &                &  $U=3$eV     & $\lambda^{\gamma,2}$  ($\lambda_C$) & 4 (251) &  &                 &  &  \\
            &                &     & $\lambda^{\epsilon,2}$ ($\lambda_D$) & -51 (10) &  &                 &  &  \\
            \midrule
YCo$_5$     & Hexagonal (I)  & DFT LSDA+U & $\lambda^{\alpha1,2}$  & -90 & $\vert\lambda^{\alpha1,2}\vert<$100$^d$ & $E(100)-E(001)$ & 365 &  567$^e$  \\
            &          SG 191      &     & $\lambda^{\alpha2,2}$  & 115 & $\vert\lambda^{\alpha 2,2}\vert<$100$^d$ &                 &  &  \\
            &                &     & $\lambda^{\gamma,2}$   & 76 &  &                 &  &  \\
            &                &     & $\lambda^{\epsilon,2}$ & 141 &  &                 &  &  \\ \midrule
Fe$_2$Si     & Trigonal (I)  & DFT GGA & $\lambda^{\alpha1,2}$  & -9 &  & $E(100)-E(001)$ & -38 &   \\
            &          SG 164      &     & $\lambda^{\alpha2,2}$  & 15 &  &                 &  &  \\
            &                &     & $\lambda^{\gamma,1}$   & 8 &  &                 &  &  \\
            &                &     & $\lambda^{\gamma,2}$ & 28 &  &                 &  &  \\ 
            &                &     & $\lambda_{12}$   & -3 &  &                 &  &  \\
            &                &     & $\lambda_{21}$ & -13 &  &                 &  &  \\
            \midrule            
L1$_0$ FePd & Tetragonal (I) & DFT GGA & $\lambda^{\alpha1,2}$  &  -21&  &         $E(100)-E(001)$ & 106 & 181$^f$ \\
            &        SG 123        &     & $\lambda^{\alpha2,2}$    & 79 &  &                 &  &  \\
            &                &     & $\lambda^{\gamma,2}$  & 31 &  &                 &  &  \\
            &                &     & $\lambda^{\epsilon,2}$   &  28 &  &                 &  &  \\
            &                &     & $\lambda^{\delta,2}$   &  106 &  &                 &  &  \\
 &  &  & $\lambda^{\alpha1,0}-\frac{\lambda^{\alpha1,2}}{3}+\frac{\lambda^{\gamma,2}}{2}$  &  & 100$^g$ &  &  &  \\\midrule 
YCo & Orthorhombic & DFT LSDA+U & $\lambda_1$  &  -11 &  &         $E(100)-E(001)$ & 22 &  \\
            &        SG 63        &     & $\lambda_2$    & 32 &  &      $E(010)-E(001)$           & -23 &  \\
            &                &     & $\lambda_3$  & 70 &  &                 &  &  \\
            &                &     & $\lambda_4$   &  -74 &  &                 &  &  \\
            &                &     & $\lambda_5$   &  -30 &  &                 &  &  \\
 &  &  & $\lambda_6$  & 7 &   &  &  &  \\
             &                &     & $\lambda_7$   &  36 &  &                 &  &  \\
            &                &     & $\lambda_8$   &  -20 &  &                 &  &  \\
 &  &  & $\lambda_9$  & 35 &   &  &  &  \\

            \bottomrule

\end{tabular}%
}
\begin{tabular}{c}
\footnotesize $^a$Ref.\cite{Handley}, $^b$Ref.\cite{Getz}, $^c$Ref.\cite{Hubert1969}, $^d$Ref.\cite{ANDREEV199559}, $^e$Ref.\cite{Nguyen_2018}, $^f$Ref.\cite{shima}, $^g$Ref.\cite{SHIMA20042173}, $^h$Ref.\cite{nieves2020spinlattice}
\end{tabular}
\end{table}

\begin{table}[H]
\centering
\caption{Elastic and magnetoelastic constants calculated using the interface between AELAS and MAELAS codes. The third column shows the DFT method used to compute the elastic constants. The experimental elastic constants of FCC Ni and BCC Fe were measured at $T\approx0$ K, while for HCP Co  $T\approx300$ K. The experimental magnetoelastic constants were estimated using the experimental elastic constants (seventh column) and the experimental magnetostrictive coefficients in Table \ref{tab:tests_MAELAS} via the relations given in Section \ref{section:theory}. The sixth column presents calculations of the elastic constants available in the Materials Project database \cite{deJong2015,Mat_Proj_1}.}
\label{tab:tests_AELAS_1}
\resizebox{\textwidth}{!}{%
\begin{tabular}{@{}ccc|cccc|ccc@{}}
\toprule
Material &
  Crystal system &
  Method &
  \begin{tabular}[c]{@{}c@{}}Elastic  \\ constant\end{tabular} &
  \begin{tabular}[c]{@{}c@{}}AELAS\\ (GPa)\end{tabular} &
  \begin{tabular}[c]{@{}c@{}}Mat.Proj.\\ (GPa)\end{tabular} &
  \begin{tabular}[c]{@{}c@{}}Expt.\\ (GPa)\end{tabular} &
  \begin{tabular}[c]{@{}c@{}}Magnetoelastic\\ constant\end{tabular}  &
  \begin{tabular}[c]{@{}c@{}}MAELAS\\ (MPa)\end{tabular} &
  \begin{tabular}[c]{@{}c@{}}Expt.\\ (MPa)\end{tabular} \\ \midrule \hline
FCC Ni      & Cubic (I)     & DFT GGA & $C_{11}$        & 298 & 276$^g$   &  261$^h$ & $b_1$ & 15.5 & 9.9  \\
            &       SG 225         &     & $C_{12}$        & 166 & 159$^g$ & 151$^h$ & $b_2$ & 19.4 &  13.9  \\
            &               &     & $C_{44}$        & 140 & 132$^g$ & 132$^h$ &  &  &     \\  & & & & & & & & & \\    
    &    & SD-MD & $C_{11}$        & 264$^f$ &    &  & $b_1$ & 10.4 &  \\
            &                &     & $C_{12}$        & 152$^f$ &  &  & $b_2$ & 14.1 &    \\
            &               &     & $C_{44}$        & 133$^f$ &  &  &  &  & \\
            \midrule
BCC Fe      & Cubic (I)     & DFT GGA & $C_{11}$        & 288 & 247$^a$   & 243$^b$  & $b_1$ & -5.2 & -4.1  \\
            &       SG 229         &     & $C_{12}$        & 152 & 150$^a$ & 138$^b$ & $b_2$ & -5.3 &  10.9  \\
            &               &     & $C_{44}$        & 104 & 97$^a$ & 122$^b$ &  &  &     \\  & & & & & & & & & \\  
      &     & SD-MD & $C_{11}$        & 230$^f$ &   &  & $b_1$ & -3.7 &  \\
            &              &     & $C_{12}$        & 134$^f$ & &  & $b_2$ & 10.6 &   \\
            &               &     & $C_{44}$        & 116$^f$ &  &  &  &  & \\
            \midrule
HCP Co      & Hexagonal (I)  & DFT LSDA+U & $C_{11}$  & 327 &   & 307$^d$ & $b_{21}$ & -21.3 & -31.9 \\
            &          SG 194      &   $J=0.8$eV   & $C_{12}$ &  157 &  &   165$^d$  &      $b_{22}$     & 48.3 & 25.5 \\
            &                &   $U=3$eV  & $C_{13}$ & 130 &  &   103$^d$  &         $b_3$  & -0.7 & -8.1 \\
            &                &     & $C_{33}$ & 308  &     &   358$^d$      & $b_4$ & 7.1 & 42.9 \\ 
           &                &     & $C_{44}$ & 69  &       &     75$^d$    &  & & \\ 
            &                &     &  &   &       &          &  &  & \\
             &                &  DFT SCAN   & $C_{11}$ & 648  &       &   &  $b_{21}$     & -51.3 &  \\
              &                &     & $C_{12}$ &  212 &   & &  $b_{22}$  &      40.5    &   \\
               &                &     & $C_{13}$ &  189 &       &  &     $b_3$   & 6.4 &  \\
                &                &     & $C_{33}$ & 633  &       &  &       $b_4$ & 8.9 &  \\
                 &                &     & $C_{44}$ &  239 &       &          &  &   & \\
                 &                &     &  &   &       &          &  &  & \\
       &  & DFT GGA & $C_{11}$  &  &  358$^c$ &  &  &  &  \\
            &                &      & $C_{12}$ &   & 165$^c$ &    &          &  & \\
            &                &     & $C_{13}$ &  & 114$^c$ &     &         &  &  \\
            &                &     & $C_{33}$ &   &  409$^c$    &       &  &  &  \\ 
           &                &     & $C_{44}$ &  &   95$^c$    &         &  & &           \\    
            \midrule
YCo$_5$     & Hexagonal (I)  & DFT GGA & $C_{11}$ & 208 & 192$^e$ & & $b_{21}$ & 14.9 &   \\
            &          SG 191      &     & $C_{12}$  & 103 & 123$^e$  &    & $b_{22}$            & -10.4 &  \\
            &                &     & $C_{13}$   & 114 & 113$^e$ &              &  $b_{3}$ & -8.0 &  \\
            &                &     & $C_{33}$ & 270 & 262$^e$ &             &  $b_{4}$  & -13.6 &  \\
              &                &     & $C_{44}$ & 49 & 48$^e$ &             &    &  &  \\             &                &     &  &   &       &          &  &  & \\
   &   & DFT LSDA+U & $C_{11}$ & -63 &  & & $b_{21}$ & 13.9 &   \\
            &               &     & $C_{12}$  & 363 &  &    & $b_{22}$            & -7.9 &  \\
            &                &     & $C_{13}$   & 115 &  &              &  $b_{3}$ & 32.5 &  \\
            &                &     & $C_{33}$ & 249 &  &             &  $b_{4}$  & -12.4 &  \\
              &                &     & $C_{44}$ & 44 & &             &    &  &  \\      
 
            \bottomrule

\end{tabular}%
}
\begin{tabular}{c}
\footnotesize $^a$Ref.\cite{Fe_MP}, $^b$Ref.\cite{Fe_elas_exp}, $^c$Ref.\cite{Co_MP}, $^d$Ref.\cite{Co_elas_exp}, $^e$Ref.\cite{YCo5_MP}, $^f$Ref.\cite{nieves2020spinlattice}, $^g$Ref.\cite{Ni_MP}, $^h$Ref.\cite{Ni_elas_exp}
\end{tabular}
\end{table}

\begin{table}[H]
\centering
\caption{Elastic and magnetoelastic constants calculated using the interface between AELAS and MAELAS codes. The third column shows the DFT method used to compute the elastic constants. The experimental elastic constants of L1$_0$ FePd were measured at $T\approx300$ K. The sixth column presents calculations of the elastic constants available in the Materials Project database \cite{deJong2015,Mat_Proj_1}. }
\label{tab:tests_AELAS_2}
\resizebox{\textwidth}{!}{%
\begin{tabular}{@{}ccc|cccc|ccc@{}}
\toprule
Material &
  Crystal system &
  Method &
  \begin{tabular}[c]{@{}c@{}}Elastic  \\ constant\end{tabular} &
  \begin{tabular}[c]{@{}c@{}}AELAS\\ (GPa)\end{tabular} &
  \begin{tabular}[c]{@{}c@{}}Mat.Proj.\\ (GPa)\end{tabular} &
  \begin{tabular}[c]{@{}c@{}}Expt.\\ (GPa)\end{tabular} &
  \begin{tabular}[c]{@{}c@{}}Magnetoelastic\\ constant\end{tabular}  &
  \begin{tabular}[c]{@{}c@{}}MAELAS\\ (MPa)\end{tabular} &
  \begin{tabular}[c]{@{}c@{}}Expt.\\ (MPa)\end{tabular} \\ \midrule \hline

Fe$_2$Si     & Trigonal (I)  & DFT GGA & $C_{11}$  & 428 & 415$^a$ &  & $b_{21}$ & 3.1 &  \\
            &          SG 164      &     & $C_{12}$  & 164 & 169$^a$ &      & $b_{22}$          & -4.2 &  \\
            &                &     &  $C_{13}$ & 133 & 133$^a$ &   & $b_{3}$             & -0.7  &  \\
            &                &     &  $C_{14}$ & -27 & -25$^a$ &   &   $b_{4}$           & 3.3 &  \\
            &                &     &  $C_{33}$ & 434 & 428$^a$ &   &    $b_{14}$          & -1.4 &  \\
                       &                &     &  $C_{44}$ &   118 & 107$^a$ &    & $b_{34}$             & -0.4 &  \\
            \midrule

L1$_0$ FePd & Tetragonal (I) & DFT GGA & $C_{11}$  & 324  & 293$^b$ &   214$^c$  &  $b_{21}$ &  -2.4 &  \\
            &        SG 123        &     &  $C_{12}$  & 67 & 62$^b$ &   143$^c$     & $b_{22}$ & -15.2 &  \\
            &                &     &  $C_{13}$ & 133 & 125$^b$ &     143$^c$    &   $b_{3}$  & -7.9 &  \\
            &                &     &  $C_{33}$  &  264 & 254$^b$ &   227$^c$ &   $b'_{3}$      & -7.9 &  \\
            &                &     & $C_{44}$   & 101  & 99$^b$ &   92$^c$   &   $b_{4}$     & -5.6 &  \\
 &  &  &  $C_{66}$ & 37 & 38$^b$ & 93$^c$ &  & &  \\
 \midrule 
YCo & Orthorhombic & DFT GGA & $C_{11}$  &  76 & 94$^d$ &  & $b_1$ & -0.9 &   \\
            &        SG 63        &     & $C_{12}$    & 45 & 61$^d$  &   &   $b_2$           &  0.6 &  \\
            &                &     & $C_{13}$  & 48 & 44$^d$ &            &  $b_3$   & -5.0 &  \\
            &                &     &  $C_{22}$ & 102  & 93$^d$ &         &   $b_4$      & 5.7 &  \\
            &                &     & $C_{23}$   & 55  & 56$^d$ &     &    $b_5$        & 0.9 &  \\
 &  &  & $C_{33}$  & 141 & 121$^d$ &  & $b_6$ & 1.5 & \\
  &  &  & $C_{44}$  & 40 & 38$^d$ &  & $b_7$  &  -5.0 &\\
   &  &  & $C_{55}$  & 27 & 29$^d$ &  & $b_8$ & 1.1 & \\
    &  &  & $C_{66}$  & 39 & 41$^d$ &  & $b_9$ & -8.2 & \\
        &  &  &   &  &  &  &  &  \\
  &  & DFT LSDA+U & $C_{11}$  &  101 &  & & $b_1$ & -1.7 &   \\
            &               &     & $C_{12}$    & 65 &  &            &   $b_2$  & 1.2 &  \\
            &                &     & $C_{13}$  & 58 &  &             &   $b_3$ & -3.8 &  \\
            &                &     &  $C_{22}$ &  94 &  &           &    $b_4$  & 4.3 &  \\
            &                &     & $C_{23}$   & 70  &  &           &   $b_5$   & -0.1 &  \\
 &  &  & $C_{33}$  & 138 &  &  & $b_6$ & 2.3 &  \\
  &  &  & $C_{44}$  & 42 &  & & $b_7$  & -4.4 &  \\
   &  &  & $C_{55}$  & 29 & & & $b_8$  &  1.1 &  \\
    &  &  & $C_{66}$  & 35 &  &  &  $b_9$ & -8.7 &  \\       
 
            \bottomrule

\end{tabular}%
}
\begin{tabular}{c}
\footnotesize $^a$Ref.\cite{Fe2Si_MP}, $^b$Ref.\cite{FePd_MP}, $^c$Ref.\cite{FePd_elas_exp}, $^d$Ref.\cite{YCo_MP}
\end{tabular}
\end{table}

\subsection{FCC Ni}

In the first example, we consider FCC Ni, which is described by Eq.\ref{eq:delta_l_cub_I} since it is a cubic (I) system. 

\subsubsection{Cell relaxation, MAE and magnetostrictive coefficients}

\begin{figure}[!ht]
\centering
\includegraphics[width=1.0\columnwidth ,angle=0]{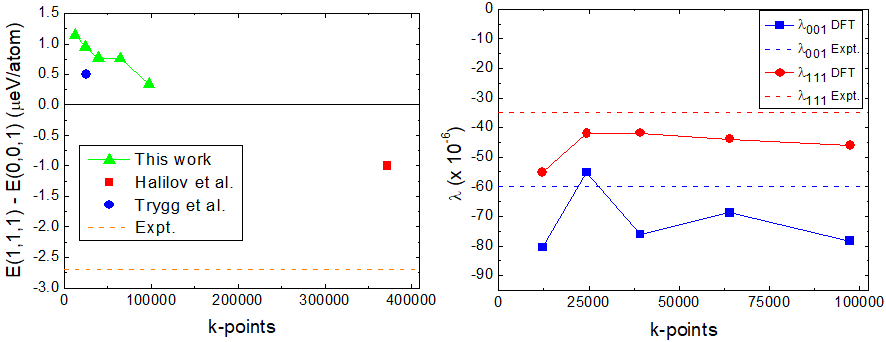}
\caption{Calculation of (left) MAE of the unstrained unit cell and (right) magnetostrictive coefficients for FCC Ni as a function of k-points.}
\label{fig:Nifcc_kp}
\end{figure}

\begin{figure}[!ht]
\centering
\includegraphics[width=1.0\columnwidth ,angle=0]{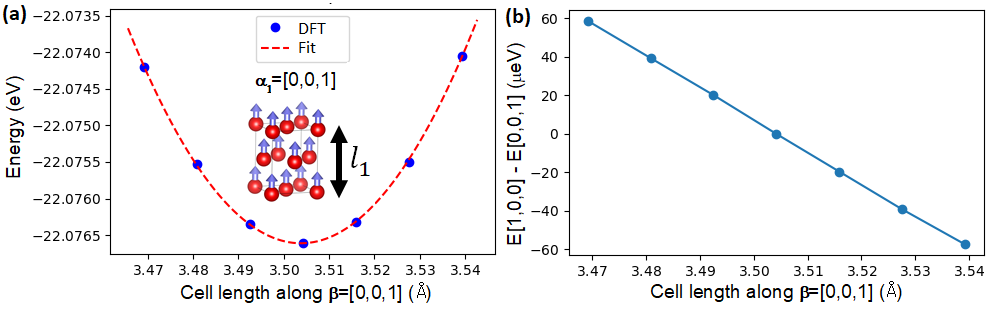}
\caption{Calculation of $\lambda_{001}$ for FCC Ni using MAELAS. (Left)  Quadratic curve fit to the energy versus cell length along $\boldsymbol{\beta}=(0,0,1)$ with spin direction $\boldsymbol{\alpha}_1=\left(0,0,1\right)$. (Right) Energy difference between states with spin directions $\boldsymbol{\alpha}_2=\left(1,0,0\right)$ and $\boldsymbol{\alpha}_1=\left(0,0,1\right)$ against the cell length along $\boldsymbol{\beta}=\left(0,0,1\right)$.}
\label{fig:Nifcc_lmb001}
\end{figure}

Firstly, we perform a cell relaxation using an automatic k-point mesh with length parameter R$_k=160$ centered on the $\Gamma$-point ($46\times46\times46$), 16 valence states, and energy cut-off $520$ eV with PAW method and GGA-PBE. The relaxed lattice parameter is $a=3.50419$ \r{A}. Next, we analyze the dependence of MAE and magnetostrictive coefficients on the k-point mesh (R$_{k}=80$, $100$, $120$, $140$ and $160$) for this relaxed unit cell using the same VASP settings as in the cell relaxation. The results are shown in Fig. \ref{fig:Nifcc_kp}. We observe that $E(1,1,1)-E(0,0,1)>0$ for all calculations up to $93150$ k-points, while the experimental value is $-2.7\mu$eV/atom \cite{kubler}. This deviation may be due to the fact that we have not used a sufficiently large number of k-points, as Halilov et al. pointed out \cite{Halilov,kubler}. Our results are in good agreement with the calculations performed by Trygg el at. where a similar number of k-points were used \cite{Trygg}. We also see that  $E(1,1,1)-E(0,0,1)$ is approaching to negative values as the number of k-points is increased. Interestingly, the calculated magnetostrictive coefficients are in quite good agreement with the experimental ones \cite{Handley} despite the deviation of  MAE for the unstrained unit cell. One possible reason for this result may be that the calculation of the magnetostrictive  coefficients involves larger energy difference between magnetization directions than the determination of MAE for the unstrained unit cell (which might be close to the accuracy limit of VASP $\sim\mu$eV), see Fig.\ref{fig:Nifcc_lmb001}. Consequently, a k-point mesh with about $10^5$ k-points may be sufficient to obtain reliable  magnetostrictive coefficients for FCC Ni using GGA, although a much more dense k-mesh would be needed to obtain a reliable  MAE for the unstrained unit cell \cite{Halilov,kubler}. Note that these two properties come from the SOC, so that in general it would be highly desirable that the used method to calculate the energies describes well both MAE and magnetostriction. Recently, we developed a spin-lattice model within the framework of coupled spin and molecular dynamics (SD-MD) that reproduces accurately the experimental elastic and magnetoelastic energies at zero-temperature \cite{nieves2020spinlattice}. We obtained very good results by applying MAELAS to this coarse-grained model of SOC, see Tables \ref{tab:tests_MAELAS} and \ref{tab:tests_AELAS_1}.

\subsubsection{Elastic and magnetoelastic constants}

To compute the elastic constants we make use of AELAS code \cite{AELAS}.  As inputs, we use the same relaxed cell and VASP settings as in the calculation of magnetostriction, but with lower number of k-points R$_k=60$ ($17\times17\times17$ for the not distorted cell) and not including SOC. Once we have the elastic constants, we use them as inputs to derive the magnetoelastic constants with MAELAS. The results are shown in Table \ref{tab:tests_AELAS_1}, where we also include calculations of elastic constants available in the Materials Project database \cite{deJong2015,Mat_Proj_1} and experimental data \cite{Ni_elas_exp}. We observe that the value of $C_{11}=298$ GPa obtained with GGA is higher than the one in the Materials Project $C_{11}=276$ GPa and in the experiment $C_{11}=261$ GPa. Concerning the magnetoelastic constants, we see that both $b_1$ and $b_2$ are in fairly good agreement with the experiment.

\subsection{BCC Fe}

In this example, we consider BCC Fe, which is described by Eq.\ref{eq:delta_l_cub_I} since it is a cubic (I) system. 

\subsubsection{Cell relaxation, MAE and magnetostrictive coefficients}

\begin{figure}[!ht]
\centering
\includegraphics[width=1.0\columnwidth ,angle=0]{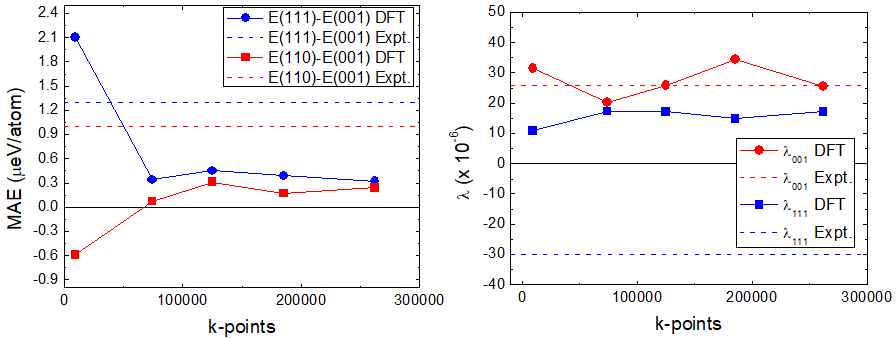}
\caption{Calculation of (left) MAE of the unstrained unit cell and (right) magnetostrictive coefficients for BCC Fe as a function of k-points.}
\label{fig:Fe_kp}
\end{figure}

\begin{figure}[!ht]
\centering
\includegraphics[width=1.0\columnwidth ,angle=0]{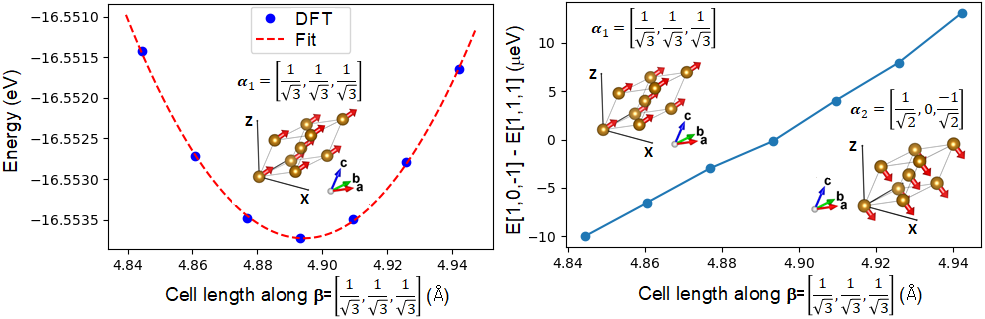}
\caption{Calculation of $\lambda_{111}$ for BCC Fe using MAELAS. (Left)  Quadratic curve fit to the energy versus cell length along $\boldsymbol{\beta}=\left(\frac{1}{\sqrt{3}},\frac{1}{\sqrt{3}},\frac{1}{\sqrt{3}}\right)$ with spin direction $\boldsymbol{\alpha}_1=\left(\frac{1}{\sqrt{3}},\frac{1}{\sqrt{3}},\frac{1}{\sqrt{3}}\right)$. (Right) Energy difference between states with spin directions $\boldsymbol{\alpha}_2=\left(\frac{1}{\sqrt{2}},0,\frac{-1}{\sqrt{2}}\right)$ and $\boldsymbol{\alpha}_1=\left(\frac{1}{\sqrt{3}},\frac{1}{\sqrt{3}},\frac{1}{\sqrt{3}}\right)$ against the cell length along $\boldsymbol{\beta}=\left(\frac{1}{\sqrt{3}},\frac{1}{\sqrt{3}},\frac{1}{\sqrt{3}}\right)$.}
\label{fig:Fe_bcc}
\end{figure}

In the first stage, we we perform a cell relaxation for the conventional cubic unit cell of the BCC (2 atoms/cell) using a $57\times57\times57$ k-mesh with 185193 k-points in the Brillouin zone. The interactions were described by a PAW potential with 14 valence electrons within the PBE approximation to the exchange-correlation, and the PW basis was generated for an energy cut-off of $380$ eV (30\% larger than the default value). The relaxed lattice parameter is $a=2.82509$ \r{A}. In Fig. \ref{fig:Fe_kp} we show the dependence of MAE and magnetostriction on the k-points for this relaxed cell using the same exchange-correlation and energy cut-off as in the cell relaxation.  The calculated values of MAE with the largest number of k-points ($262144
$) are $E(110)-E(001)=0.32\mu$eV/atom and $E(111)-E(001)=0.24\mu$eV/atom which are a bit lower than the experimental values $1\mu$eV/atom and $1.34\mu$eV/atom, respectively \cite{Getz}. Concerning the magnetostrictive coefficients, we obtained $\lambda_{001}=25.7\times10^{-6}$ and $\lambda_{111}=17.2\times10^{-6}$, while the experimental values at $T=4.2$K are $\lambda_{001}=26\times10^{-6}$ and $\lambda_{111}=-30\times10^{-6}$ \cite{Handley}. We see that $\lambda_{001}$ is quite close to the experimental result, while  $\lambda_{111}$ is in good agreement with previous DFT calculations \cite{Zhang2012,Fahnle2002,Burkert} but it has the opposite sign as the experimental value.  The calculation of $\lambda_{111}$ is presented in Fig.\ref{fig:Fe_bcc}. We observe that the sign of the derivative of the energy difference between states with spin directions $\boldsymbol{\alpha}_2=\left(\frac{1}{\sqrt{2}},0,\frac{-1}{\sqrt{2}}\right)$ and $\boldsymbol{\alpha}_1=\left(\frac{1}{\sqrt{3}},\frac{1}{\sqrt{3}},\frac{1}{\sqrt{3}}\right)$  with respect to the cell length along  $\boldsymbol{\beta}=\left(\frac{1}{\sqrt{3}},\frac{1}{\sqrt{3}},\frac{1}{\sqrt{3}}\right)$ evaluated at $l=l_2$ is equal to the sign of the calculated $\lambda_{111}$ ($>0$), as expected from Eq.\ref{eq:lambda_derivative}. However, the experimental  $\lambda_{111}$ is negative. This deviation might be due to a possible failure of GGA related to the location of the Fermi level in a region of majority band t2g density of states  \cite{Jones2015,guo2002}. Aiming to clarify the influence of MAELAS in this result, we applied MAELAS to a spin-lattice model for BCC Fe, that reproduces accurately the experimental elastic and magnetoelastic energies, obtaining almost the same experimentally observed magnetostriction \cite{nieves2020spinlattice}, see Tables \ref{tab:tests_MAELAS} and \ref{tab:tests_AELAS_1}.
 
\subsubsection{Elastic and magnetoelastic constants}

As inputs for AELAS code \cite{AELAS}, we use the same relaxed cell and VASP settings as in the calculation of magnetostriction, but with lower number of k-points R$_k=60$ ($21\times21\times21$ for the not distorted cell) and not including SOC. Once we have the elastic constants, we use them as inputs to derive the magnetoelastic constants with MAELAS. The results are shown in Table \ref{tab:tests_AELAS_1}, where we also include calculations of elastic constants available in the Materials Project database \cite{deJong2015,Mat_Proj_1} and experimental data \cite{Fe_elas_exp}. We observe that the value of $C_{11}=288$ GPa obtained with AELAS is significantly higher than the one in the Materials Project $C_{11}=247$ GPa and in the experiment $C_{11}=243$ GPa. Regarding the magnetoelastic constants, we see that the value for $b_1=-5.2$ MPa generated with MAELAS is close to the estimated experimental value $b_1=-4.1$ MPa. However, we obtain a negative sign for $b_2=-5.4$ MPa, while in the experiment it is positive $b_2=10.9$ MPa. This deviation is due to the positive sign of $\lambda_{111}$ given by DFT that we have mentioned above, see Eq.\ref{eq:lamb_cub} \cite{Zhang2012,Jones2015,guo2002}.

\subsection{HCP Co}

As a first example of hexagonal (I) system, we consider HCP Co.  

\subsubsection{Cell relaxation, MAE and magnetostrictive coefficients}

\begin{figure}[!ht]
\centering
\includegraphics[width=1.0\columnwidth ,angle=0]{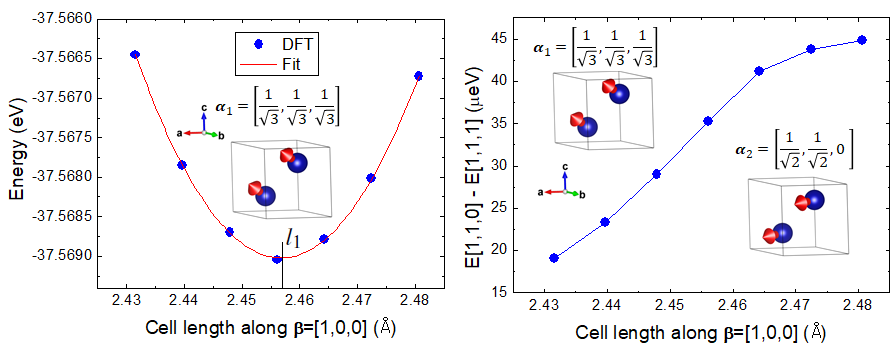}
\caption{Calculation of $\lambda^{\alpha1,2}$ for HCP Co using MAELAS with the meta-GGA functional SCAN. (Left)  Quadratic curve fit to the energy versus cell length along $\boldsymbol{\beta}=\left(1,0,0\right)$ with spin direction $\boldsymbol{\alpha}_1=\left(\frac{1}{\sqrt{3}},\frac{1}{\sqrt{3}},\frac{1}{\sqrt{3}}\right)$. (Right) Energy difference between states with spin directions $\boldsymbol{\alpha}_2=\left(\frac{1}{\sqrt{2}},\frac{1}{\sqrt{2}},0\right)$ and $\boldsymbol{\alpha}_1=\left(\frac{1}{\sqrt{3}},\frac{1}{\sqrt{3}},\frac{1}{\sqrt{3}}\right)$ against the cell length along $\boldsymbol{\beta}=\left(1,0,0\right)$.}
\label{fig:Co_hcp_L1}
\end{figure}

For this material, we set the length parameter R$_k=160$ for the generation of the automatic k-point mesh, which for the relaxed (not distorted) cell, results in a 75$\times$75$\times$40 k-point grid with 250000 points in the Brillouin zone. All calculations were done with an energy cut-off $406.563$ eV (50\% larger than the default one), 15 electrons in the valence states, and the meta-GGA functional SCAN \cite{scan}, with aspherical contributions to the PAW one-centre terms. The relaxed lattice parameters are $a=b=2.4561$ \r{A} and $c=3.9821$ \r{A}. The calculated MAE for the relaxed cell is $E(100)-E(001)=53\mu$eV/atom which is quite close to the experimental value $61\mu$eV/atom \cite{Getz}. As it is shown in Table \ref{tab:tests_MAELAS}, the calculated magnetostrictive coefficients are also close to the experimental ones, except for $\lambda^{\epsilon,2}$. Similarly, converting them into Mason's definitions via the relations given by Eq.\ref{eq:lamb_Mason_hex_I}, we see that only $\lambda_D=-1\times10^{-6}$ is significantly deviated from the experiment ($-128\times10^{-6}$) \cite{Hubert1969}. Aiming to clarify this result, we performed a direct calculation of $\lambda_D$ using $\boldsymbol{\beta}=\left(\frac{1}{\sqrt{2}},0,\frac{1}{\sqrt{2}}\right)$, $\boldsymbol{\alpha}_1=\left(\frac{1}{\sqrt{2}},0,\frac{1}{\sqrt{2}}\right)$ and $\boldsymbol{\alpha}_2=\left(0,0,1\right)$ finding $\lambda_D=-9\times10^{-6}$, which is consistent with the indirect calculation through Clark's definition but still far from the experimental value.  Fig.\ref{fig:Co_hcp_L1} shows the quadratic curve fit to the energy versus cell length along $\boldsymbol{\beta}=(1,0,0)$  with $\boldsymbol{\alpha}_1=\left(\frac{1}{\sqrt{3}},\frac{1}{\sqrt{3}},\frac{1}{\sqrt{3}}\right)$ to calculate $\lambda^{\alpha1,2}$.

\begin{figure}[!ht]
\centering
\includegraphics[width=1.0\columnwidth ,angle=0]{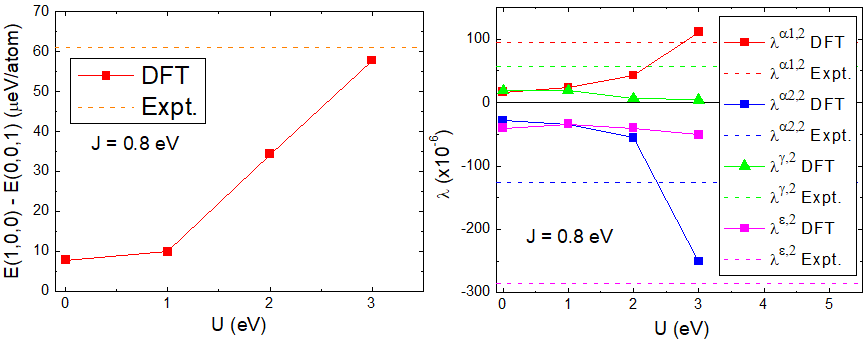}
\caption{Calculation of (left) MAE of the unstrained unit cell and (right) magnetostrictive coefficients for HCP Co using the LSDA+U approach with different values of parameter $U$.}
\label{fig:Co_hcp_U}
\end{figure}

We have also performed a second test using the rotationally invariant LSDA+U approach introduced by Liechtenstein et al. \cite{LSDA_Lie} fixing $J=0.8$eV and varying $U$ on the d-electrons \cite{Nguyen_2018}. In all calculations we use the same pseudopotential and number of k-points as in the tests performed with SCAN. The energy cut-off is set to 380 eV. In this case the relaxed lattice parameters are $a=b=2.48896$ \r{A} and $c=4.02347$ \r{A}. The analysis of MAE and magnetostriction for different values of $U$ is shown in Fig.\ref{fig:Co_hcp_U}. We see that MAE approximates the experimental value at $U=3$eV, so that we might expect a reliable description of SOC for this value of $U$. Increasing $U$ up to $3$eV has a significant effect on $\lambda^{\alpha1,2}$ and $\lambda^{\alpha2,2}$ making them to approach the experimental values. On the other hand, $\lambda^{\gamma,2}$ and $\lambda^{\epsilon,2}$ don't change too much within the range of values used for $U$. The sign of all magnetostrictive coefficients are in good agreement to the experimental ones. However, as in the case with SCAN, $\lambda^{\epsilon,2}$ is significantly underestimated. Possible reasons for this systematic deviation might be a failure of DFT \cite{guo2002}, the applied deformations (we used the default value $0.01$ for the tag $-s$ that sets the maximum value of parameter $s$ in the generation of the deformed unit cells, see Eq.\ref{eq:strain_hex_I}), the used VASP settings (k-point mesh, exchange-correlation functional, smearing method, lattice parameters, ...) or higher order corrections in the equation of the relative length change  Eq.\ref{eq:delta_l_hex_I} \cite{Mishima}. This issue should be further investigated to clarify its possible causes.

\subsubsection{Elastic and magnetoelastic constants}

For the calculation of the elastic constants we use the same relaxed cell and VASP settings as in the calculation of magnetostriction, but without SOC and lower number of k-points R$_k=60$ ($28\times28\times15$ for the not distorted cell). In addition to SCAN, we also run calculations with LSDA+U setting $J=0.8eV$ and $U=3$eV. In Table \ref{tab:tests_AELAS_1}, we see that LSDA+U and GGA (Materials Project database \cite{Ni_MP}) give better results than SCAN for both the elastic and magnetoelastic constants. The magnetoelastic constants obtained with LSDA+U are moderately good, except for $b_3$ and $b_4$ which are one order of magnitude lower than in the experiment, mainly due to the deviations coming from $\lambda^{\gamma,2}$ and $\lambda^{\epsilon,2}$ given by MAELAS, see Table \ref{tab:tests_MAELAS}.

\subsection{YCo$_5$}

In this example we study the hexagonal (I) system YCo$_5$  with prototype CaCu$_5$ structure (space group 191). 

\subsubsection{Cell relaxation, MAE and magnetostrictive coefficients}

\begin{figure}[!ht]
\centering
\includegraphics[width=1.0\columnwidth ,angle=0]{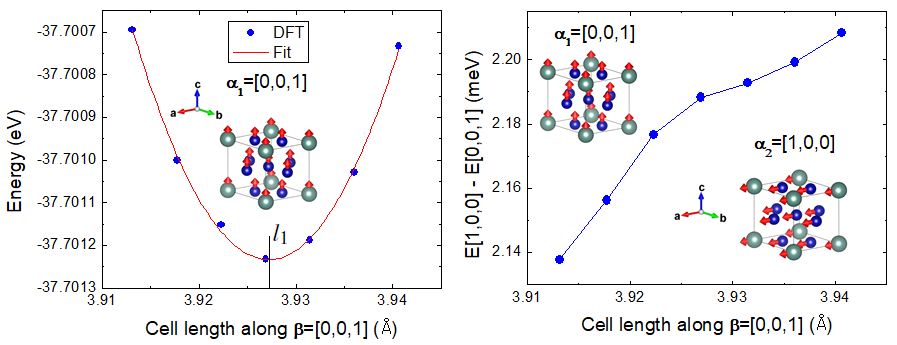}
\caption{Calculation of $\lambda^{\alpha2,2}$ for YCo$_5$ using MAELAS. (Left)  Quadratic curve fit to the energy versus cell length along $\boldsymbol{\beta}=\left(0,0,1\right)$ with spin direction $\boldsymbol{\alpha}_1=\left(0,0,1\right)$. (Right) Energy difference between states with spin directions $\boldsymbol{\alpha}_2=\left(1,0,0\right)$ and $\boldsymbol{\alpha}_1=\left(0,0,1\right)$ against the cell length along $\boldsymbol{\beta}=\left(0,0,1\right)$.}
\label{fig:YCo5_L2}
\end{figure}

We use the simplified (rotationally invariant) approach to the LSDA+U introduced by Dudarev et al. \cite{LSDA} with parameters $U=1.9$ eV and $J=0.8$ eV for Co, and $U=J=0$ eV for Y given in Ref.\cite{Nguyen_2018}. For the calculation of the relaxed cell, MAE and magnetostrictive coefficients we used an automatic k-point mesh with length parameter R$_k=100$ centered on the $\Gamma$-point ($23\times23\times25$ for the not distorted cell), 11 and 9 valence states for Y and Co, respectively, and energy cut-off $375$ eV. The cell relaxation leads to lattice parameters $a=b=4.9253$ \r{A} and $c=3.9269$ \r{A}. The calculated MAE is $E(100)-E(001)=365 \mu$eV/atom which is lower than the experimental value $567 \mu$eV/atom \cite{Nguyen_2018}. Andreev measured the magnetostriction along a and c axis, finding that the magnitude of $\vert\lambda^{\alpha1,2}\vert$ and $\vert\lambda^{\alpha2,2}\vert$ can not be greater than $10^{-4}$ \cite{ANDREEV199559}. We obtained $\lambda^{\alpha1,2}=-90\times10^{-6}$ and $\lambda^{\alpha2,2}=115\times10^{-6}$ which are quite close to the experimental upper limit. In Fig.\ref{fig:YCo5_L2} we present the quadratic curve fit to the energy versus cell length along $\boldsymbol{\beta}=(0,0,1)$ with $\boldsymbol{\alpha}_1=\left(0,0,1\right)$ to calculate $\lambda^{\alpha2,2}$.

\subsubsection{Elastic and magnetoelastic constants}

The calculation of the elastic constants is performed using the same relaxed cell and VASP settings as for magnetostriction, but without SOC and lower number of k-points R$_k=60$ ($14\times14\times15$ for the not distorted cell). In addition to LSDA+U, we also run calculations with GGA. In Table \ref{tab:tests_AELAS_1}, we observe that LSDA+U leads to an unstable phase ($C_{11}-C_{12}<0$), while GGA gives better results.

\subsection{Fe$_2$Si}

To illustrate the application of MAELAS to trigonal (I) systems, we apply it to Fe$_2$Si (space group 164) \cite{Fe2Si_exp}.

\subsubsection{Cell relaxation, MAE and magnetostrictive coefficients}

For the calculation of the cell relaxation, MAE and magnetostrictive coefficients we used an automatic k-point mesh with length parameter R$_k=80$ centered on the $\Gamma$-point ($24\times24\times17$ for the not distorted cell), 14 and 4 valence states for Fe and Si, respectively, and energy cut-off $520$ eV with PAW method and GGA-PBE. The relaxed lattice parameters are $a=3.9249$ \r{A} and $c=4.8311$ \r{A}. The calculated MAE is $E(100)-E(001)=-38 \mu$eV/atom (easy plane). Sun et al. reported MAE values with the screened hybrid Heyd-Scuseria-Ernzerhof (HSE06) functional smaller than with PBE for 2D Fe$_2$Si \cite{Fe2Si_dft,HSE06}. Chi Pui Tang et al. calculated some electronic properties for bulk Fe$_2$Si finding that the  densities  of states  in the  vicinity of  the Fermi  level  is mainly contributed from the d-electrons of Fe \cite{Fe2Si_dft_3D}. In Table \ref{tab:tests_MAELAS}, we observe that the overall anisotropic magnetostriction given by MAELAS is rather small, which makes this material interesting for high-flux core applications because it can reduce hysteresis loss \cite{Fe2Si_app}.
\begin{figure}[!ht]
\centering
\includegraphics[width=1.0\columnwidth ,angle=0]{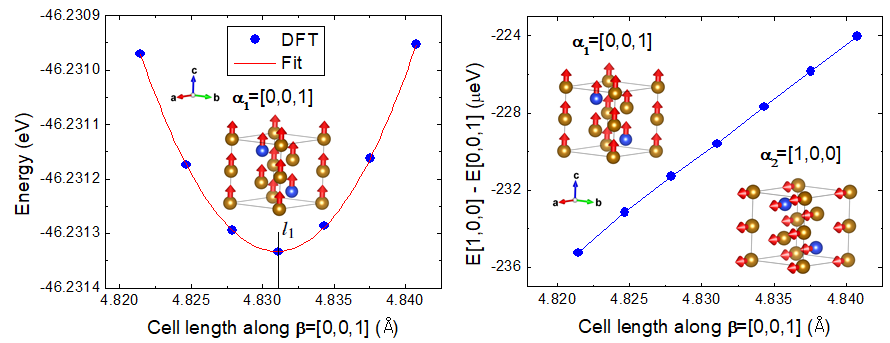}
\caption{Calculation of $\lambda^{\alpha2,2}$ for Fe$_2$Si using MAELAS. (Left)  Quadratic curve fit to the energy versus cell length along $\boldsymbol{\beta}=\left(0,0,1\right)$ with spin direction $\boldsymbol{\alpha}_1=\left(0,0,1\right)$. (Right) Energy difference between states with spin directions $\boldsymbol{\alpha}_2=\left(1,0,0\right)$ and $\boldsymbol{\alpha}_1=\left(0,0,1\right)$ against the cell length along $\boldsymbol{\beta}=\left(0,0,1\right)$.}
\label{fig:Fe2Si_L2}
\end{figure}

\subsubsection{Elastic and magnetoelastic constants}

As inputs for AELAS, we use the same relaxed cell and VASP settings as in the calculation of magnetostriction, but without SOC and lower number of k-points R$_k=60$ ($18\times18\times12$ for the not distorted cell). In Table \ref{tab:tests_AELAS_2}, we see that AELAS gives similar elastic constants as in the Materials Project \cite{Fe2Si_MP}. The derived magnetoelastic constants are small which is consistent with the low magnetostrictive coefficients that we obtained previously.

\subsection{L1$_0$ FePd}

As an example of tetragonal (I) system, we calculate the anisotropic magnetostrictive coefficients of L1$_0$ FePd (space group 123).  

\subsubsection{Cell relaxation, MAE and magnetostrictive coefficients}

\begin{figure}[!ht]
\centering
\includegraphics[width=1.0\columnwidth ,angle=0]{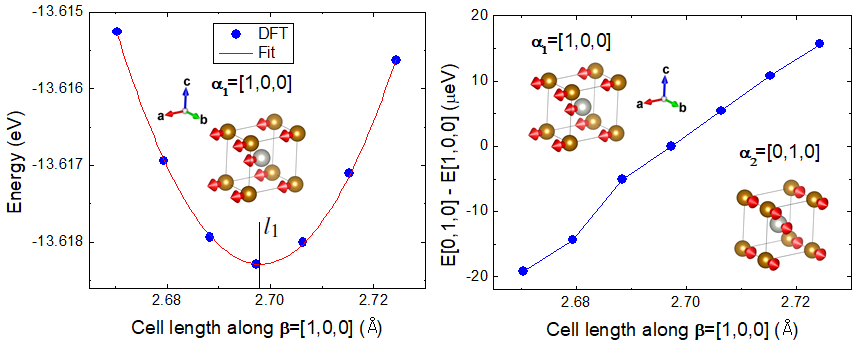}
\caption{Calculation of $\lambda^{\gamma,2}$ for L1$_0$ FePd using MAELAS. (Left)  Quadratic curve fit to the energy versus cell length along $\boldsymbol{\beta}=\left(1,0,0\right)$ with spin direction $\boldsymbol{\alpha}_1=\left(1,0,0\right)$. (Right) Energy difference between states with spin directions $\boldsymbol{\alpha}_2=\left(0,1,0\right)$ and $\boldsymbol{\alpha}_1=\left(1,0,0\right)$ versus the cell length along $\boldsymbol{\beta}=\left(1,0,0\right)$.}
\label{fig:FePd_L3}
\end{figure}

For the calculation of the cell relaxation, MAE and magnetostrictive coefficients we used an automatic k-point mesh with length parameter R$_k=100$ centered on the $\Gamma$-point ($37\times37\times27$ for the not distorted cell), 8 and 10 valence states for Fe and Pd, respectively, and energy cut-off $375$ eV with PAW method and GGA-PBE. The relaxed lattice parameters are $a=2.6973$ \r{A} and $c=3.7593$ \r{A}. We obtained a MAE $E(100)-E(001)=106 \mu$eV/atom which is lower than in the experiment $181 \mu$eV/atom \cite{shima}. The values of the obtained anisotropic magnetostrictive coefficients are shown in Table \ref{tab:tests_MAELAS}. Shima et al. reported a relative length change equal to $100\times10^{-6}$ along a-axis under a magnetic field in the same direction ($\boldsymbol{\beta}=\boldsymbol{\alpha}=(1,0,0)$)  \cite{SHIMA20042173}. According to Eq.\ref{eq:delta_l_tet_I}, this measurement corresponds to  $\lambda^{\alpha1,0}-\frac{\lambda^{\alpha1,2}}{3}+\frac{\lambda^{\gamma,2}}{2}$. For the anisotropic part of this quantity, we obtained $-\frac{\lambda^{\alpha1,2}}{3}+\frac{\lambda^{\gamma,2}}{2}=22.5\times10^{-6}$. In Fig.\ref{fig:FePd_L3} we show the quadratic curve fit to the energy versus cell length along $\boldsymbol{\beta}=(1,0,0)$  with $\boldsymbol{\alpha}_1=\left(1,0,0\right)$ to calculate $\lambda^{\gamma,2}$.

\subsubsection{Elastic and magnetoelastic constants}

The calculation of the elastic constants with AELAS is performed using the same relaxed cell and VASP settings as for magnetostriction, but without SOC and lower number of k-points R$_k=60$ ($22\times22\times16$ for the not distorted cell). As we see in Table \ref{tab:tests_AELAS_2}, we obtain similar results as in the Materials Project database \cite{FePd_MP}.

\subsection{YCo}

For the case of orthorhombic systems, we study the compound YCo (space group 63) \cite{YCo_exp}.

\subsubsection{Cell relaxation, MAE and magnetostrictive coefficients}

In this case, we use the simplified (rotationally invariant) approach to the LSDA+U \cite{LSDA} with parameters $U=1.9$ eV and $J=0.8$ eV for Co, and $U=J=0$ eV for Y in the same way as in YCo$_5$ \cite{Nguyen_2018}. For the calculation of the relaxed cell, MAE and magnetostrictive coefficients we used an automatic k-point mesh with length parameter R$_k=90$ centered on the $\Gamma$-point ($22\times9\times23$ for the not distorted cell), 11 and 9 valence states for Y and Co, respectively, and energy cut-off $375$ eV. The cell relaxation leads to lattice parameters $a=4.0686$ \r{A}, $b=10.3157$ \r{A} and $c=3.8957$ \r{A}. As we see in Table \ref{tab:tests_MAELAS}, both MAE and magnetostriction are quite small for this material.
\begin{figure}[!ht]
\centering
\includegraphics[width=1.0\columnwidth ,angle=0]{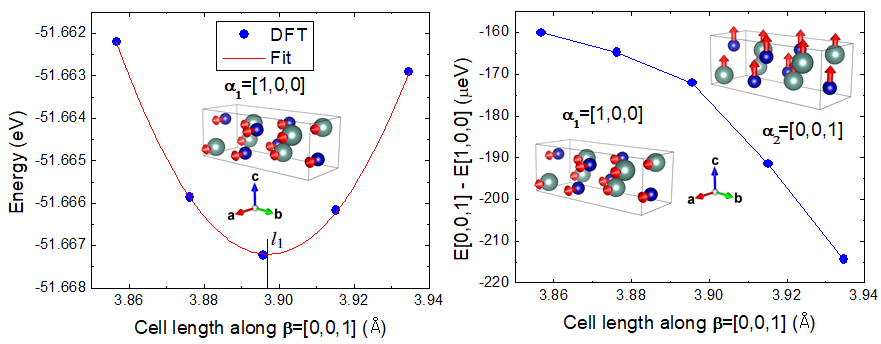}
\caption{Calculation of $\lambda_5$ for YCo using MAELAS. (Left)  Quadratic curve fit to the energy versus cell length along $\boldsymbol{\beta}=\left(0,0,1\right)$ with spin direction $\boldsymbol{\alpha}_1=\left(1,0,0\right)$. (Right) Energy difference between states with spin directions $\boldsymbol{\alpha}_2=\left(0,0,1\right)$ and $\boldsymbol{\alpha}_1=\left(1,0,0\right)$ against the cell length along $\boldsymbol{\beta}=\left(0,0,1\right)$.}
\label{fig:YCo_L5}
\end{figure}

\subsubsection{Elastic and magnetoelastic constants}

The calculation of the elastic constants is performed using the same relaxed cell and VASP settings as for magnetostriction, but without SOC and lower number of k-points R$_k=60$ ($15\times6\times15$ for the not distorted cell). In addition to LSDA+U, we also run calculations with GGA. In Table \ref{tab:tests_AELAS_2}, we observe that both LSDA+U and GGA lead to similar elastic and magnetoelastic constants.

\section{Conclusions and future perspectives}
\label{section:conclusion}

In summary, the program MAELAS offers computational tools to tackle the complex phenomenon of magnetostriction by automated first-principles calculations. It could potentially be used to discover and design novel magnetostrictive materials by a high-throughput screening approach. In particular, materials with giant magnetostriction (beyond conventional cubic and hexagonal systems), isotropic or very low magnetostriction (like FeNi alloys) might be of technological importance.

The preliminary tests of the program show quite encouraging results, although there is still room for improvement. First principle calculations are still quite challenging for materials with very low MAE or localized 4f-electrons \cite{Nguyen_2018}. In this sense, MAELAS could also be a useful tool to understand, test and improve the DFT methods to compute induced properties by SOC and crystal field interactions like the MAE of unstrained systems and anisotropic magnetostriction.

Presently, we are working on new features of MAELAS and online visualization tools \cite{maelasviewer}. We are also considering to increase the number of supported crystal systems, as well as to implement more computationally efficient methods to calculate magnetoelastic constants and magnetostrictive coefficients. These extensions might be included in new versions of the code.

\section*{Acknowledgement}

This work was supported by the ERDF in the IT4Innovations national supercomputing center - path to exascale project (CZ.02.1.01/0.0/0.0/16-013/0001791) within the OPRDE.
This work was supported by The Ministry of Education, Youth and Sports from  the Large Infrastructures for Research, Experimental Development, and Innovations project “e-INFRA CZ - LM2018140”. This work was supported by the Donau project No. 8X20050 and the computational resources provided by the Open Access Grant Competition of IT4Innovations National Supercomputing Center within the projects OPEN-18-5, OPEN-18-33, and OPEN-19-14. DL, SA, and APK acknowledge the Czech Science Foundations grant No.~20-18392S.  P.N., D.L., and~S.A. acknowledge support from the H2020-FETOPEN no.~863155 s-NEBULA project.

\appendix

\section{Conversion between different definitions of magnetostrictive coefficients for hexagonal (I)}
\label{app:hex_I}

The magnetostrictive coefficients for hexagonal (I) shown in Eq.\ref{eq:delta_l_hex_I} were defined by Clark et al. in 1965 \cite{Clark}. However, one can find other definitions like those given by Mason \cite{Mason}, Birss \cite{Birss}, and Callen and Callen \cite{Callen}. In this appendix, we show the conversion formulas between these definitions and those provided by Clark et al. \cite{Clark} (Eq.\ref{eq:delta_l_hex_I}).

\subsection{Mason's form}

In 1954, based on a general thermodynamic function with stresses and intensity of magnetization as the fundamental variables \cite{Mason1951}, Mason derived the following form of the relative length change \cite{Mason}
\begin{equation}
\begin{aligned}
     \frac{\Delta l}{l_0}\Bigg\vert_{\boldsymbol{\beta}}^{\boldsymbol{\alpha}} & =\lambda^{\alpha1,0}_{Mason}(\beta_x^2+\beta_y^2)+\lambda^{\alpha2,0}_{Mason}\beta_z^2+\lambda_A[(\alpha_x\beta_x+\alpha_y\beta_y)^2-(\alpha_x\beta_x+\alpha_y\beta_y)\alpha_z\beta_z]\\
     & + \lambda_B[(1-\alpha_z^2)(1-\beta_z^2)-(\alpha_x\beta_x+\alpha_y\beta_y)^2]\\
     & +\lambda_C[(1-\alpha_z^2)\beta_z^2-(\alpha_x\beta_x+\alpha_y\beta_y)\alpha_z\beta_z] + 4\lambda_D(\alpha_x\beta_x+\alpha_y\beta_y)\alpha_z\beta_z.
    \label{eq:delta_l_Mason_hex_I}
\end{aligned}
\end{equation}
These magnetostrictive coefficients are related to those defined in Eq.\ref{eq:delta_l_hex_I} as \cite{Clark}
\begin{equation}
\begin{aligned}
     \lambda^{\alpha1,0}_{Mason} & = \lambda^{\alpha1,0}+\frac{2}{3}\lambda^{\alpha1,2}\\
     \lambda^{\alpha2,0}_{Mason} & =  \lambda^{\alpha2,0}+\frac{2}{3}\lambda^{\alpha2,2}\\
     \lambda_A & =  -\lambda^{\alpha1,2}+\frac{1}{2}\lambda^{\gamma,2}\\
     \lambda_B & =  -\lambda^{\alpha1,2}-\frac{1}{2}\lambda^{\gamma,2}\\
     \lambda_C & =  -\lambda^{\alpha2,2}\\
     \lambda_D & = \frac{1}{2} \lambda^{\epsilon,2}-\frac{1}{4}\lambda^{\alpha1,2}+\frac{1}{8}\lambda^{\gamma,2}-\frac{1}{4}\lambda^{\alpha2,2}.
    \label{eq:lamb_Mason_hex_I}
\end{aligned}
\end{equation}
Note in the original work of Mason \cite{Mason} the terms that describes the volume magnetostriction were not included. Here we added these terms ($\lambda^{\alpha1,0}_{Mason}$, $\lambda^{\alpha2,0}_{Mason}$) in order to fully recover the Eq.\ref{eq:delta_l_hex_I}.

\subsection{Birss's form}

In 1959 Birss derived an equivalent equation of relative length change in this form \cite{Birss}
\begin{equation}
\begin{aligned}
     \frac{\Delta l}{l_0}\Bigg\vert_{\boldsymbol{\beta}}^{\boldsymbol{\alpha}} & =Q_0+Q_1\beta_z^2+Q_2(1-\alpha_z^2)+Q_4(1-\alpha_z^2)\beta_z^2+Q_6(\alpha_x\beta_x+\alpha_y\beta_y)\alpha_z\beta_z\\
     & + Q_8(\alpha_x\beta_x+\alpha_y\beta_y)^2.
    \label{eq:delta_l_Birss_hex_I}
\end{aligned}
\end{equation}
These magnetostrictive coefficients are related to those defined in Eq.\ref{eq:delta_l_hex_I} as \cite{Clark}
\begin{equation}
\begin{aligned}
     Q_0 & = \lambda^{\alpha1,0}+\frac{2}{3}\lambda^{\alpha1,2}\\
     Q_1 & =  \lambda^{\alpha2,0}+\frac{2}{3}\lambda^{\alpha2,2}-\lambda^{\alpha1,0}-\frac{2}{3}\lambda^{\alpha1,2}\\
     Q_2 & =  -\lambda^{\alpha1,2}-\frac{1}{2}\lambda^{\gamma,2}\\
     Q_4 & =  \lambda^{\alpha1,2}+\frac{1}{2}\lambda^{\gamma,2}-\lambda^{\alpha2,2}\\
     Q_6 & =  2\lambda^{\epsilon,2}\\
     Q_8 & = \lambda^{\gamma,2}.
    \label{eq:lamb_Birss_hex_I}
\end{aligned}
\end{equation}

\subsection{Callen and Callen's form}

In 1965 Callen and Callen obtained other equivalent form of the equation of relative length change by including two-ion interactions into the theory of magnetostriction arising from single-ion crystal-field effects \cite{Callen}. It reads
\begin{equation}
\begin{aligned}
     \frac{\Delta l}{l_0}\Bigg\vert_{\boldsymbol{\beta}}^{\boldsymbol{\alpha}} & =\frac{1}{3}\lambda_{11}^\alpha+\frac{1}{2\sqrt{3}}\lambda_{12}^{\alpha}\left(\alpha_z^2-\frac{1}{3}\right)+2\lambda_{21}^\alpha\left(\beta_z^2-\frac{1}{3}\right)\\
     & +\sqrt{3}\lambda_{22}^\alpha\left(\alpha_z^2-\frac{1}{3}\right)\left(\beta_z^2-\frac{1}{3}\right)+\lambda^{\gamma}\left[\frac{1}{2}(\alpha_x^2-\alpha_y^2)(\beta_x^2-\beta_y^2)+2\alpha_x\alpha_y\beta_x\beta_y\right]\\
     & + 2\lambda^{\epsilon}(\alpha_x\alpha_z\beta_x\beta_z+\alpha_y\alpha_z\beta_y\beta_z),
    \label{eq:delta_l_Callen_hex_I}
\end{aligned}
\end{equation}
These magnetostrictive coefficients are related to those defined in Eq.\ref{eq:delta_l_hex_I} as \cite{Callen}
\begin{equation}
\begin{aligned}
     \lambda_{11}^\alpha & = 2\lambda^{\alpha1,0}+\lambda^{\alpha2,0}+2\lambda^{\alpha1,2}+\lambda^{\alpha2,2}\\
     \lambda_{12}^\alpha & =  \frac{4}{\sqrt{3}}\lambda^{\alpha1,2}+\frac{2}{\sqrt{3}}\lambda^{\alpha2,2}\\
    \lambda_{21}^\alpha & = -\frac{1}{2} \lambda^{\alpha1,0}+\frac{1}{2}\lambda^{\alpha2,0}\\
     \lambda_{22}^\alpha & =  -\frac{1}{\sqrt{3}}\lambda^{\alpha1,2}+\frac{1}{\sqrt{3}}\lambda^{\alpha2,2}\\
     \lambda^{\epsilon} & =  \lambda^{\epsilon,2}\\
     \lambda^{\gamma} & = \lambda^{\gamma,2}.
    \label{eq:lamb_Callen_hex_I}
\end{aligned}
\end{equation}

\section{Conversion between different definitions of magnetostrictive coefficients for tetragonal (I)}
\label{app_tet_I}

In 1994 Cullen et al. \cite{Cullen} derived the equation of relative length change given by  Eq.\ref{eq:delta_l_tet_I} for tetragonal (I) system. In 1954 Mason obtained an equivalent equation that reads \cite{Mason}
\begin{equation}
\begin{aligned}
     \frac{\Delta l}{l_0}\Bigg\vert_{\boldsymbol{\beta}}^{\boldsymbol{\alpha}} & =\lambda^{\alpha1,0}_{Mason}(\beta_x^2+\beta_y^2) +\lambda^{\alpha2,0}_{Mason}\beta_z^2 +\frac{1}{2}\lambda_1[(\alpha_x\beta_x-\alpha_y\beta_y)^2-(\alpha_x\beta_y+\alpha_y\beta_x)^2\\
     & +(1-\beta_z^2)(1-\alpha_z^2)-2\alpha_z\beta_z(\alpha_x\beta_x+\alpha_y\beta_y)]+4\lambda_2\alpha_z\beta_z(\alpha_x\beta_x+\alpha_y\beta_y)\\
     & +4\lambda_3\alpha_x\alpha_y\beta_x\beta_y+\lambda_4[\beta_z^2(1-\alpha_z^2)-\alpha_z\beta_z(\alpha_x\beta_x+\alpha_y\beta_y)]\\
     & +\frac{1}{2}\lambda_5[(\alpha_x\beta_y-\alpha_y\beta_x)^2-(\alpha_x\beta_x+\alpha_y\beta_y)^2+(1-\beta_z^2)(1-\alpha_z^2)].
    \label{eq:delta_l_Mason_tet_I}
\end{aligned}
\end{equation}
These magnetostrictive coefficients are related to those defined by Eq.\ref{eq:delta_l_tet_I} in the following way 
\begin{equation}
    \begin{aligned}
        \lambda^{\alpha1,0}_{Mason} & = \lambda^{\alpha1,0}+\frac{2}{3}\lambda^{\alpha1,2}\\
     \lambda^{\alpha2,0}_{Mason} & =  \lambda^{\alpha2,0}+\frac{2}{3}\lambda^{\alpha2,2}\\
        \lambda_{1} & = -\lambda^{\alpha1,2}+\frac{1}{2}\lambda^{\gamma,2} \\
        \lambda_{2} & = \frac{1}{2}\lambda^{\epsilon,2}-\frac{1}{4}\lambda^{\alpha2,2}-\frac{1}{4}\lambda^{\alpha1,2}+\frac{1}{8}\lambda^{\gamma,2} \\
        \lambda_{3} & = \frac{1}{2}\lambda^{\delta,2}-\lambda^{\alpha1,2} \\
        \lambda_{4} & = -\lambda^{\alpha2,2} \\
        \lambda_{5} & = -\lambda^{\alpha1,2}-\frac{1}{2}\lambda^{\gamma,2}.
    \label{eq:lamb_tet_Mason}
    \end{aligned}
\end{equation}
Note in the original work of Mason \cite{Mason} the terms that describes the volume magnetostriction were not included. Here we added these terms ($\lambda^{\alpha1,0}_{Mason}$, $\lambda^{\alpha2,0}_{Mason}$) in order to fully recover the Eq.\ref{eq:delta_l_tet_I}.

\section{Generation of the deformed unit cells}
\label{app_matrix}

In this appendix we present the procedure to generate the deformed unit cells for the calculation of each magnetostrictive coefficient.  The deformed unit cells are generated by multiplying the lattice vectors of the initial unit cell $\boldsymbol{a}=(a_x,a_y,a_z)$, $\boldsymbol{b}=(b_x,b_y,b_z)$, $\boldsymbol{c}=(c_x,c_y,c_z)$ by the deformation gradient $F_{ij}$  \cite{Tadmor_2009}
\begin{equation}
\begin{pmatrix}
a'_{x} & b'_{x} & c'_{x}\\
a'_{y} & b'_{y} & c'_{y}\\
a'_{z} & b'_{z} & c'_{z} \\
\end{pmatrix}
= 
\begin{pmatrix}
F_{xx} & F_{xy} & F_{xz}\\
F_{yx} & F_{yy} & F_{yz}\\
F_{zx} & F_{zy} & F_{zz} \\
\end{pmatrix}\cdot
\begin{pmatrix}
a_{x} & b_{x} & c_{x}\\
a_{y} & b_{y} & c_{y}\\
a_{z} & b_{z} & c_{z} \\
\end{pmatrix}
\label{eq:deform_latt0}
\end{equation}
where $a'_i$, $b'_i$ and $c'_i$ ($i=x,y,z$) are the components of the lattice vectors of the deformed cell. In the infinitesimal strain theory, the deformation gradient is related to the displacement gradient ($\partial u_i / \partial r_j$) as $F_{ij}=\delta_{ij}+ \partial u_i/\partial r_j$, where $\delta_{ij}$ is the Kronecker delta. Hence, according to Eq.\ref{eq:stiffnees_tensor}, the strain tensor $\epsilon_{ij}$ can be written in terms of the deformation gradient as 
\begin{equation}
\boldsymbol{\epsilon}=
\begin{pmatrix}
\epsilon_{xx} & \epsilon_{xy} & \epsilon_{xz}\\
\epsilon_{yx} & \epsilon_{yy} & \epsilon_{yz}\\
\epsilon_{zx} & \epsilon_{zy} & \epsilon_{zz} \\
\end{pmatrix}
= 
\frac{1}{2}\begin{pmatrix}
2(F_{xx}-1) & F_{xy}+F_{yx} & F_{xz}+F_{zx} \\
F_{xy}+F_{yx}  & 2(F_{yy}-1) & F_{yz}+F_{zy} \\
F_{xz}+F_{zx}  & F_{yz}+F_{zy}  & 2(F_{zz}-1) 
\end{pmatrix}
.
\label{eq:strain_deform0}
\end{equation}
 In MAELAS, we consider deformation gradients to optimize the unit cell in the measuring directions $\boldsymbol{\beta}$ given by Table \ref{tab:beta_alpha_data}. Additionally, we also constrain the determinant of the deformation gradients to be equal to one ($det(\boldsymbol{F})= 1$) in order to preserve the volume of the unit cells (isochoric deformation) \cite{guo2000,Burkert}. We point out that there are other possible variants of the following deformation modes  \cite{guo2002}.

\subsection{Cubic (I) system}
\label{subsection:cubic_deform}

For cubic (I) systems MAELAS generates two set of deformed unit cells with tetragonal deformations along $\boldsymbol{\beta}=(0,0,1)$ and trigonal deformations along $\boldsymbol{\beta}=(1/\sqrt{3},1/\sqrt{3},1/\sqrt{3})$ to calculate $\lambda_{001}$ and $\lambda_{111}$, respectively (see Table \ref{tab:beta_alpha_data}). The deformation gradients for these two deformation modes are
\begin{equation}
\boldsymbol{F}_{\boldsymbol{\beta}=(0,0,1)}^{\lambda_{001}}(s)=
\begin{pmatrix}
\frac{1}{\sqrt{1+s}} & 0 & 0\\
0 & \frac{1}{\sqrt{1+s}} & 0\\
0 & 0 & 1+s \\
\end{pmatrix}
,\quad
\boldsymbol{F}_{\boldsymbol{\beta}=\left(\frac{1}{\sqrt{3}},\frac{1}{\sqrt{3}},\frac{1}{\sqrt{3}}\right)}^{\lambda_{111}}(s)=\zeta
\begin{pmatrix}
1 & \frac{s}{2} & \frac{s}{2}\\
\frac{s}{2} & 1 & \frac{s}{2}\\
\frac{s}{2} & \frac{s}{2} & 1 \\
\end{pmatrix}
\label{eq:strain_cub_I}
\end{equation}
where $\zeta=\sqrt[3]{4/(4-3s^2+s^3)}$. The parameter $s$ controls the applied deformation, and its maximum value can be specified through the command line of the program MAELAS using tag $-s$. The total number of deformed cells can be chosen with tag $-n$.

\subsection{Hexagonal (I) system}

\begin{figure}[ht!]
\centering
\includegraphics[width=0.18\columnwidth ,angle=0]{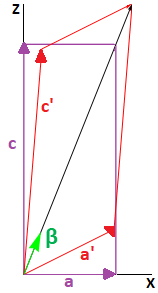}
\caption{Sketch of the deformation generated by the deformation gradient $\boldsymbol{F}^{\lambda^{\epsilon,2}}$ given by Eq.\ref{eq:strain_hex_I} to calculate $\lambda^{\epsilon,2}$. The purple line represents the relaxed cell with lattice parameters $(a,b,c)$, while the red line stands for the deformed cell with lattice parameters $(a',b',c')$. This deformation meets the property $\boldsymbol{\beta}=\frac{\boldsymbol{a}+\boldsymbol{c}}{\vert\boldsymbol{a}+\boldsymbol{c}\vert}=\frac{\boldsymbol{a'}+\boldsymbol{c'}}{\vert\boldsymbol{a'}+\boldsymbol{c'}\vert}$.}
\label{fig:deform_lam_eps}
\end{figure}

In the case of hexagonal (I), MAELAS generates 4 sets of deformed cells using the following deformation gradients
\begin{equation}
\begin{aligned}
\boldsymbol{F}\Big\vert_{\boldsymbol{\beta}=(1,0,0)}^{\lambda^{\alpha1,2}} (s) & =
\begin{pmatrix}
1+s & 0 & 0\\
0 & \frac{1}{\sqrt{1+s}} & 0\\
0 & 0 & \frac{1}{\sqrt{1+s}} \\
\end{pmatrix}
,
\boldsymbol{F}\Big\vert_{\boldsymbol{\beta}=(0,0,1)}^{\lambda^{\alpha2,2}}(s)=\begin{pmatrix}
\frac{1}{\sqrt{1+s}} & 0 & 0\\
0 & \frac{1}{\sqrt{1+s}} & 0\\
0 & 0 & 1+s \\
\end{pmatrix} \\
\boldsymbol{F}\Big\vert_{\boldsymbol{\beta}=(1,0,0)}^{\lambda^{\gamma,2}}(s)& =
\begin{pmatrix}
1+s & 0 & 0\\
0 & \frac{1}{\sqrt{1+s}} & 0\\
0 & 0 & \frac{1}{\sqrt{1+s}} \\
\end{pmatrix}
,
\boldsymbol{F}\Big\vert_{\boldsymbol{\beta}=\frac{\left(a,0,c\right)}{\sqrt{a^2+c^2}}}^{\lambda^{\epsilon,2}} (s)=\omega
\begin{pmatrix}
1 & 0 & \frac{s c}{2a}\\
0 & 1 & 0\\
\frac{s a}{2c} & 0& 1 \\
\end{pmatrix}
\label{eq:strain_hex_I}
\end{aligned}
\end{equation}
where $\omega=\sqrt[3]{4/(4-s^2)}$, $a$ and $c$ are the lattice parameters of the relaxed (not deformed) unit cell. The fractions $c/a$ and $a/c$ were introduced in the deformation gradient elements $F_{xz}^{\lambda^{\epsilon,2}}$ and $F_{zx}^{\lambda^{\epsilon,2}}$, respectively, in order to generate deformations that meet the property $\boldsymbol{\beta}=\frac{\boldsymbol{a}+\boldsymbol{c}}{\vert\boldsymbol{a}+\boldsymbol{c}\vert}=\frac{\boldsymbol{a'}+\boldsymbol{c'}}{\vert\boldsymbol{a'}+\boldsymbol{c'}\vert}$, where $\boldsymbol{a'}$ and $\boldsymbol{c'}$ are the lattice vectors of the distorted unit cell, see Fig. \ref{fig:deform_lam_eps}. This deformation mode makes it easy to compute the cell length $l$ in the measuring direction $\boldsymbol{\beta}$ since it is just $l=\vert\boldsymbol{a'}+\boldsymbol{c'}\vert$. It is inspired by the trigonal deformation for the cubic (I) case $\boldsymbol{F}^{\lambda_{111}}$ where deformations meet the property $\boldsymbol{\beta}=\frac{\boldsymbol{a}+\boldsymbol{b}+\boldsymbol{c}}{\vert\boldsymbol{a}+\boldsymbol{b}+\boldsymbol{c}\vert}=\frac{\boldsymbol{a'}+\boldsymbol{b'}+\boldsymbol{c'}}{\vert\boldsymbol{a'}+\boldsymbol{b'}+\boldsymbol{c'}\vert}$. 

\subsection{Trigonal (I) system}

In the case of trigonal (I), MAELAS generates 6 sets of deformed unit cells using the following deformation gradients
\begin{equation}
\begin{aligned}
 & \boldsymbol{F}\Big\vert_{\boldsymbol{\beta}=(1,0,0)}^{\lambda^{\alpha1,2}}  (s) =
\begin{pmatrix}
1+s & 0 & 0\\
0 & \frac{1}{\sqrt{1+s}} & 0\\
0 & 0 & \frac{1}{\sqrt{1+s}} \\
\end{pmatrix}
,
\boldsymbol{F}\Big\vert_{\boldsymbol{\beta}=(0,0,1)}^{\lambda^{\alpha2,2}} (s) =\begin{pmatrix}
\frac{1}{\sqrt{1+s}} & 0 & 0\\
0 & \frac{1}{\sqrt{1+s}} & 0\\
0 & 0 & 1+s \\
\end{pmatrix} \\
& \boldsymbol{F}\Big\vert_{\boldsymbol{\beta}=(1,0,0)}^{\lambda^{\gamma,1}} (s)=
\begin{pmatrix}
1+s & 0 & 0\\
0 & \frac{1}{\sqrt{1+s}} & 0\\
0 & 0 & \frac{1}{\sqrt{1+s}} \\
\end{pmatrix}\\
& \boldsymbol{F}\Big\vert_{\boldsymbol{\beta}=\frac{\left(a,0,c\right)}{\sqrt{a^2+c^2}}}^{\lambda^{\gamma,2}}(s)=\boldsymbol{F}\Big\vert_{\boldsymbol{\beta}=\frac{\left(a,0,c\right)}{\sqrt{a^2+c^2}}}^{\lambda_{12}}(s)=\boldsymbol{F}\Big\vert_{\boldsymbol{\beta}=\frac{\left(a,0,c\right)}{\sqrt{a^2+c^2}}}^{\lambda_{21}}(s)=\omega
\begin{pmatrix}
1 & 0 & \frac{s c}{2a}\\
0 & 1 & 0\\
\frac{s a}{2c} & 0& 1 \\
\end{pmatrix}.
\label{eq:strain_trig_I}
\end{aligned}
\end{equation}

\subsection{Tetragonal (I) system}

In the case of tetragonal (I), MAELAS generates 5 sets of deformed cells using the following deformation gradients
\begin{equation}
\begin{aligned}
& \boldsymbol{F}\Big\vert_{\boldsymbol{\beta}=(1,0,0)}^{\lambda^{\alpha1,2}} (s)  =
\begin{pmatrix}
1+s & 0 & 0\\
0 & \frac{1}{\sqrt{1+s}} & 0\\
0 & 0 & \frac{1}{\sqrt{1+s}} \\
\end{pmatrix}
,
\boldsymbol{F}\Big\vert_{\boldsymbol{\beta}=(0,0,1)}^{\lambda^{\alpha2,2}}(s)=\begin{pmatrix}
\frac{1}{\sqrt{1+s}} & 0 & 0\\
0 & \frac{1}{\sqrt{1+s}} & 0\\
0 & 0 & 1+s \\
\end{pmatrix} \\
& \boldsymbol{F}\Big\vert_{\boldsymbol{\beta}=(1,0,0)}^{\lambda^{\gamma,2}}(s) =
\begin{pmatrix}
1+s & 0 & 0\\
0 & \frac{1}{\sqrt{1+s}} & 0\\
0 & 0 & \frac{1}{\sqrt{1+s}} \\
\end{pmatrix}
,
\boldsymbol{F}\Big\vert_{\boldsymbol{\beta}=\frac{\left(a,0,c\right)}{\sqrt{a^2+c^2}}}^{\lambda^{\epsilon,2}} (s)=\omega
\begin{pmatrix}
1 & 0 & \frac{s c}{2a}\\
0 & 1 & 0\\
\frac{s a}{2c} & 0& 1 \\
\end{pmatrix}\\
& \boldsymbol{F}\Big\vert_{\boldsymbol{\beta}=\left(\frac{1}{\sqrt{2}},\frac{1}{\sqrt{2}},0\right)}^{\lambda^{\delta,2}}  (s)=
\omega
\begin{pmatrix}
1 & \frac{s}{2} & 0\\
\frac{s}{2} & 1 & 0\\
0 & 0 & 1 
\end{pmatrix}.
\label{eq:strain_tet_I}
\end{aligned}
\end{equation}

\subsection{Orthorhombic system}

For orthorhombic crystals MAELAS generates 9 sets of deformed cells using the following deformation gradients
\begin{equation}
\begin{aligned}
 & \boldsymbol{F}\Big\vert_{\boldsymbol{\beta}=(1,0,0)}^{\lambda_1} (s) =
\begin{pmatrix}
1+s & 0 & 0\\
0 & \frac{1}{\sqrt{1+s}} & 0\\
0 & 0 & \frac{1}{\sqrt{1+s}} \\
\end{pmatrix}
,
\boldsymbol{F}\Big\vert_{\boldsymbol{\beta}=(1,0,0)}^{\lambda_2} (s) =
\begin{pmatrix}
1+s & 0 & 0\\
0 & \frac{1}{\sqrt{1+s}} & 0\\
0 & 0 & \frac{1}{\sqrt{1+s}} \\
\end{pmatrix} \\
 & \boldsymbol{F}\Big\vert_{\boldsymbol{\beta}=(0,1,0)}^{\lambda_3} (s) =
\begin{pmatrix}
 \frac{1}{\sqrt{1+s}} & 0 & 0\\
0 & 1+s & 0\\
0 & 0 & \frac{1}{\sqrt{1+s}} \\
\end{pmatrix}
,
\boldsymbol{F}\Big\vert_{\boldsymbol{\beta}=(0,1,0)}^{\lambda_4} (s) =
\begin{pmatrix}
 \frac{1}{\sqrt{1+s}} & 0 & 0\\
0 & 1+s & 0\\
0 & 0 & \frac{1}{\sqrt{1+s}} \\
\end{pmatrix} \\
 & \boldsymbol{F}\Big\vert_{\boldsymbol{\beta}=(0,0,1)}^{\lambda_5} (s) =
\begin{pmatrix}
 \frac{1}{\sqrt{1+s}} & 0 & 0\\
0 & \frac{1}{\sqrt{1+s}} & 0\\
0 & 0 & 1+s \\
\end{pmatrix}
,
\boldsymbol{F}\Big\vert_{\boldsymbol{\beta}=(0,0,1)}^{\lambda_6} (s) =
\begin{pmatrix}
 \frac{1}{\sqrt{1+s}} & 0 & 0\\
0 & \frac{1}{\sqrt{1+s}} & 0\\
0 & 0 & 1+s \\
\end{pmatrix} \\
& \boldsymbol{F}\Big\vert_{\boldsymbol{\beta}=\frac{\left(a,b,0\right)}{\sqrt{a^2+b^2}}}^{\lambda_7} (s) =\omega
\begin{pmatrix}
1 & \frac{s b}{2a} & 0\\
\frac{s a}{2b} & 1 & 0\\
0 & 0 & 1 \\
\end{pmatrix}
,
\boldsymbol{F}\Big\vert_{\boldsymbol{\beta}=\frac{\left(a,0,c\right)}{\sqrt{a^2+c^2}}}^{\lambda_8} (s) =\omega
\begin{pmatrix}
1 & 0 & \frac{s c}{2a}\\
0 & 1 & 0\\
\frac{s a}{2c} & 0 & 1 \\
\end{pmatrix}\\
& \boldsymbol{F}\Big\vert_{\boldsymbol{\beta}=\frac{\left(0,b,c\right)}{\sqrt{b^2+c^2}}}^{\lambda_9} (s) =\omega
\begin{pmatrix}
1 & 0 & 0\\
0 & 1 & \frac{s c}{2b}\\
0 & \frac{s b}{2c} & 1 \\
\end{pmatrix}.
\label{eq:strain_ortho}
\end{aligned}
\end{equation}






\bibliographystyle{elsarticle-num}
\bibliography{mybibfile.bib}







\end{document}